\documentclass[times,11pt]{article}
\usepackage{amsmath}
\usepackage{amsfonts}
\usepackage{graphicx}
\usepackage{float}
\usepackage{sectsty}
\usepackage{natbib}
\numberwithin{equation}{section}
\usepackage[toc,page]{appendix}
\usepackage{algorithm2e}
\usepackage{hyperref}
\usepackage{abstract}
\usepackage{pgfplots}
\usepackage{geometry,comment,dsfont}

\usepackage{caption,subcaption}
\usepackage{lscape}
 \rm

\begin{document}
\newgeometry{left=2.5cm,right=2.5cm,bottom=2cm,top=2cm}
\title{An Introduction to Applications of Wavelet Benchmarking with Seasonal Adjustment}
%\author{H,J,D,H}
\author{Homesh Sayal\\University of Cambridge \and John A. D.
Aston\thanks{Address for correspondance: John Aston, Statistical Laboratory,
DPMMS, University of Cambridge, CB3 0WB, \texttt{j.aston@statslab.cam.ac.uk}}
\\University of Cambridge \and Duncan Elliott\\Office for National Statistics
\and Hernando Ombao\\University of California, Irvine}
%\affiliation{$^1$ University of Cambridge, $^2$ Office for National Statistics,
%$^3$ University of California, Irvine}
\date{\today}
\maketitle
\begin{abstract}
\hspace{5mm}Prior to adjustment, accounting conditions between national accounts
data sets are frequently violated. Benchmarking is the procedure used by economic
agencies to make such data sets consistent. It typically involves adjusting a
high frequency time series (e.g. quarterly data) so it becomes consistent with a
lower frequency version (e.g. annual data). Various methods have been developed
to approach this problem of inconsistency between data sets. This paper
introduces a new statistical procedure; namely wavelet benchmarking. Wavelet
properties allow high and low frequency processes to be jointly analysed and we
show that benchmarking can be formulated and approached succinctly in the
wavelet domain. Furthermore the time and frequency localisation properties of
wavelets are ideal for handling more complicated benchmarking problems. The
versatility of the procedure is demonstrated using simulation studies where we
provide evidence showing it substantially outperforms currently used methods. Finally, we apply this novel method of wavelet benchmarking to official Office of National Statistics (ONS) data.

\end{abstract}
%\tableofcontents

\section{Introduction}
National Statistics Institues (NSIs) such as the Office for National Statistics (ONS) are
responsible for collecting and analysing economic data e.g. national accounts
data and labour data \cite[Ch~1]{DagumCholetteBook(2006)}. Data sets collected by such agencies are typically adjusted for a variety of reasons. Benchmarking (the focus of this paper) is an adjustment procedure \cite[Ch~1]{DagumCholetteBook(2006)} used to make measurements from the same statistical process across different periodicities consistent. Since
national accounts data must satisfy specific accounting conditions, benchmarking
has important applications. It is well documented for example that unmodified
quarterly GDP data is not consistent with its annual GDP version (i.e. the
quarterly totals do not sum to the corresponding annual value) \cite[Ch~1]{DagumCholetteBook(2006)}. Since
data sets of different periodicities are collected from different sample surveys
and compiled differently such discrepancies occur naturally as a result of survey errors. In many cases, for
example, a larger sample is used for the less frequent survey; hence the lower
frequency series is typically more reliable than its corresponding high
frequency version. The aim of benchmarking is to adjust the high frequency
series so it becomes consistent with the lower frequency version while
preserving short term fluctuations. The low frequency and adjusted high
frequency time series are referred to as the benchmark and benchmarked series
respectively.

Benchmarking can be considered as a subclass of signal extraction problems.  Current literature can be classified as providing either numerical or model
based solutions. Denton \cite{Denton(71)} approached benchmarking using a numerical method
based on quadratic minimisation. A penalty function defined by the user
specifies this minimisation procedure. Dagum and Cholette \cite{DagumCholette(94)} expressed
benchmarking in terms of a stochastic regression model; hence a regression type
solution is provided. The Denton method is computationally simple but sometimes yields poor solutions. Dagum and Cholette's method often requires fitting
complex structural time series models; this creates the
problem of estimating ARIMA components which can be difficult using traditional
methods. However, since it has a regression setting, confidence intervals can be
obtained and so uncertainty about point estimates can be quantified. In
practice, NSIs often implement methods which make
simplifying assumptions to allow for easier estimation and greater transparency of the model.

In this paper, we present a new non-parametric methodology for benchmarking. It is based on the natural idea that the time series can be decomposed into different
time-scale components, and these components used to constrain the high
frequency series. Wavelets \cite{Daubechies(92)} provide a natural time-frequency
decomposition, and are able to adapt to local conditions in the time series. This is important in macro-economic times series routinely analysed by the ONS. Wavelets extend the ideas of Fourier
decompositions by removing the assumption of stationarity in the time series. By
combining data sets from different wavelet decomposition levels, and making use of
the unbalanced Haar decomposition \cite{Fryzlewicz(2007)} to account for the non-dyadic nature
of the analysis, our proposed method is able to reconstruct a benchmarked series with high
frequency components that still satisfy the low frequency constraints.

Outliers and abrupt structural changes are commonplace in observed time series. Current methods provide global benchmarking solutions; hence volatile regions of the high frequency series have the potential to introduce artefacts into the benchmarked series. The time-frequency localisation properties of wavelets \cite*[Page~59]{PercivalWalden(2000)} provide a local solution to benchmarking and thus overcome such a problem. While wavelet bases considered in this paper only depend of the length of observed time series, bases dependent on the structure of the observed time series can easily be constructed if required.

In addition, NSIs frequently publish a seasonally adjusted version of the high frequency series. Seasonal adjustment is another procedure applied to data in order to remove unwanted effects \cite{findley2005}, but care has to be taken when combining seasonal adjustment and benchmarking.
Along with adjustments for calendar effects (e.g. trading day effects) a version of benchmarking must be applied so both the original and seasonally adjusted high frequency series satisfy the benchmark constraint. We show that by using a suitable seasonal model, wavelet benchmarking and seasonal adjustment can be combined within the same framework.

The paper proceeds as follows. Section \ref{Background} provides an introduction to current benchmarking methods and a short introduction to wavelets. Section \ref{Methodology} describes the process of benchmarking in the wavelet domain. Additional issues which require consideration such as thresholding and seasonal adjustment are also discussed. In Section \ref{dataanalysissection} wavelet benchmarking is applied to a variety of simulated data and official ONS data. Section \ref{Conclusion} concludes the paper. Details on the simulation implementations are given in the Appendix.

\section{Background}\label{Background}
The requirement of benchmarking is frequently demanded by the ONS. Currently a variety of benchmarking methods are proposed in Denton \cite{Denton(71)}, Dagum and Cholette \cite{DagumCholette(94)} and Durbin and Quenneville \cite{DurbinQuenneville(97)} to name but a few. Many official statistics at the ONS are currently benchmarked using a Denton approach but a methodological shift to Dagum and Cholette benchmarking has been recommended \cite{BrownPS2012} and is being introduced into production systems. We will therefore consider these two approaches and provide a comparison of wavelet benchmarking to them.

Consider the following introductory example. A quarterly GDP time series needs to be
benchmarked to an annual GDP time series; typically the annual series is less
noisy than its quarterly version. To simplify the benchmarking procedure many
NSIs assume such an annual series is not contaminated with noise
(binding benchmarking). Throughout this paper the above example is used to provide a concrete description of benchmarking, however the methodology is applicable to general periodicity relations. For completeness the following expresses benchmarking in a more formal way.

Suppose the stochastic process $Y^{H}_{T}(t)$ describes the true evolution of a high (i.e. quarterly) frequency process. $Y^{H}_{O}(t)$, $Y^{L}_{O}(t)$ denote high and low (i.e. annual) frequency observed versions of a statistical process respectively. The disturbance terms $\epsilon^{H}(t)$, $\epsilon^{L}(t)$ contaminate the true processes. In a discrete time setting:
\begin{align*}
& Y^{H}_{O,t}=Y^{H}_{T,t}+\epsilon^{H}_{t}, \hspace{2mm} t=1,\ldots, n \\
& Y^{L}_{O,s}=g_{s}(Y^{H}_{T})+\epsilon^{L}_{s}, \hspace{2mm} s=1,\ldots, m
\end{align*}
where $g_{s}()$ represents some function of the underlying series, often a summation over a small range, and $Y^{H}_{T}=\left[ Y^{H}_{T,1},\ldots,Y^{H}_{T,n}\right]$. In the setting of quarterly to annual benchmarking, $g_{s}(Y^{H}_{T})=\sum\limits_{i=4s-3}^{4s}Y^{H}_{T,i}$, and $n=4m$. While subsequent methods rely upon various statistical techniques they have a fundamental similarity in how benchmarking may be interpreted. The estimated series $(\hat{Y}^{H}_{T})$ can typically be expressed as a linear combination of the observed high $(Y^{H}_{O})$ and low $(Y^{L}_{O})$ frequency processes. This results in the following estimator:
\begin{align}
\label{benchmarkingsolutiongeneral}
\hat{Y}^{H}_{T}
=
A
\left(\begin{array}{c}
Y^{H}_{O} \\
Y^{L}_{O} \\
\end{array}\right),
\end{align}
where $\hat{Y}_{T}^{H}=\left[\hat{Y}_{T,1}^{H},\ldots,\hat{Y}_{T,n}^{H}
\right]^{'},$
$Y^{H}_{O}=\left[Y_{O,1}^{H},\ldots,Y_{O,n}^{H}\right]^{'}$,
$Y^{L}_{O}=\left[Y_{O,1}^{L},\ldots,Y_{O,m}^{L}\right]^{'}$.

Embedded within matrix $A$ is information describing the relationship between
the high $(Y^{H}_{O})$ and low frequency series $(Y^{L}_{O})$. Conditional
on the benchmarking procedure implemented, additional information summarising
statistical features, such as the time series correlation structure or estimates of model parameters may be present. In particular for the parametric (non parametric) approach, the matrix $A$ is explicitly (implicitly) data dependent.

\subsection{Denton Method}\label{denton1}
Denton benchmarking \cite{Denton(71)}, the first widely used benchmarking procedure, is based on the principle of movement preservation. This ensures that the benchmarked
high frequency series $\hat{Y}^{H}_{T}$ evolves similarly to the
observed series $Y^{H}_{O}$ (i.e. $\hat{Y}^{H}_{T}$ is approximately a level shift or proportionate to $Y^{H}_{O}$ depending on which Denton method is implemented). As described in \cite*[Chapter~6]{DagumCholetteBook(2006)} the Denton method has the
following underlying model for discrete data:
\begin{align}
\label{Dentonbencmarking1}& Y^{H}_{O,t}=Y^{H}_{T,t}+\epsilon_{t}, \\
\label{Dentonbenchmarking2}&
Y^{L}_{O,s}=\sum\limits_{t=p_{s,1}}^{p_{s,k}}j_{s,t}Y^{H}_{T,t}
\end{align}
In quarterly to annual binding benchmarking $j_{s,t}=1$, with $p_{s,1}$ and $p_{s,4}$ representing the beginning and end quarters corresponding to year $s$ respectively.

Two primary variants of Denton benchmarking are additive and proportional differencing with each best suited for additive and multiplicative time series respectively. Additive first differencing  keeps the discrepancy between the benchmarked and original series $\hat{Y}^{H}_{T,t}-Y^{H}_{O,t}$ as close as possible to a constant by minimising the following objective function (equation \ref{additiveobjectivefunction1}) subject to the benchmark constraint (equation \ref{additiveobjectivefunction1a}) being satisfied:
\begin{align}
&\left(Y^{H}_{T,1}-Y^{H}_{O,1}\right) +\sum\limits_{t=2}^{n} \left[
\left(Y^{H}_{T,t}-Y^{H}_{O,t}\right)
-\left(Y^{H}_{T,t-1}-Y^{H}_{O,t-1}\right)\right]^{2} \label{additiveobjectivefunction1},\text{ subject to} \\
&Y^{L}_{O,s}=\sum\limits_{t=t_{s,1}}^{t_{s,k}}j_{s,t}Y^{H}_{T,t}, \hspace{2mm}\forall s=1,\ldots,m \label{additiveobjectivefunction1a}
\end{align}
The benchmarked series is approximately a vertical shift of the original series, i.e. $\hat{Y}^{H}_{T,t}\approx Y^{H}_{O,t}+c$, $c\in \mathbb{R},$ $\forall t$.

Denton \cite{Denton(71)} devised the following solution based on Lagrangian \cite*[Chapter~6]{BinmoreDavies(2002)} optimisation:
\begin{align}
\label{Dentonsolution}
\hat{Y}^{H}_{T}=Y^{H}_{O}+C\left[Y^{L}_{O}-B^{'}Y^{H}_{O}\right],
\hspace{2mm}C=A^{-1}B\left(B^{'}A^{-1}B\right)^{-1}
\end{align}
\begin{align}
\label{DentonMatrix1}
B=
\left(\begin{array}{cccc}
\mathbf{j} & \mathbf{0} & \ldots & \mathbf{0} \\
\mathbf{0} & \mathbf{j} & \ldots & \mathbf{0} \\
\vdots & \vdots &  & \vdots \\
\mathbf{0} & \mathbf{0} & \ldots & \mathbf{j} \\
\end{array}\right)_{n\times m},
\hspace{5mm}
A=D^{'}D
\hspace{5mm}\text{with}\hspace{5mm}
D=
\left(\begin{array}{cccccc}
1&0&0&\ldots&0&0 \\
\hspace{-2.5mm}-1&1&0&\ldots&0&0\\
0&\hspace{-2.5mm}-1&1&\ldots&0&0\\
\vdots&\vdots&\vdots&&\vdots&\vdots\\
0&0&0&\ldots&\hspace{-2.5mm}-1&1\\
\end{array}\right)_{n\times n}
\end{align}
where $\mathbf{j}$ and $\mathbf{0}$ are $l=\frac{n}{m}$ dimensional column vectors taking values one and zero respectively. In the aforementioned example, $\mathbf{j}=[1\hspace{1mm} 1\hspace{1mm} 1\hspace{1mm} 1\hspace{1mm}]^{'}$, $B^{'}$ annualises the quarterly series, and $D$ calculates the differences between the error terms $\epsilon_{t}$. Equation \ref{Dentonsolution} expresses the benchmarked series as a linear combination of the noisy quarterly series and non noisy annual series. The following expresses the solution in the form of equation \ref{benchmarkingsolutiongeneral}:
\begin{align}
\label{Dentonsolutiongernalform}
\hat{Y}^{H}_{T}
=
\left(\begin{array}{cc}
I-CB^{'} & C \\
\end{array}\right)
\left(\begin{array}{c}
Y^{H}_{O} \\
Y^{L}_{O} \\
\end{array}\right)
\end{align}
It is possible to specify equation \ref{additiveobjectivefunction1} in terms of higher order additive differences between the original and adjusted series. For example $\left(\sum\limits_{i=1}^{n}(\Delta^{h}
\hat{Y}_{T,t}^{H}-\Delta^{h} Y^{H}_{O,t})^{2}\right)$ corresponds to the $h^{th}$
order additive model with $\Delta^{h}$ being the $h^{th}$ difference operator and
values outside the adjustment range being defined as
$Y^{H}_{O,t}=Y^{H}_{T,t}, \hspace{2mm} t=0,-1,\ldots,1-h$. A solution is provided to this example by setting $A=\underbrace{D^{'}\ldots D^{'}}_{h}\underbrace{D\ldots D}_{h}$ in equation \ref{Dentonsolution}. The data analysis section implements additive Denton benchmarking with values of $h=1,2$. Hereafter such benchmarked series are referred to as Denton 1 and Denton 2 respectively.

While not being computationally demanding and only requiring basic assumptions on the structural form of the time series being analysed, the Denton method occasionally performs poorly. This is evident in time series which evolve unconventionally; for example consider a time series containing a small number of extreme data points. In this case, the Denton method would adjust a disproportionate number of data points. As mentioned in the Introduction, one motivation for considering wavelets is that their time-frequency localisation properties can help overcome this problem.

\subsection{Dagum and Cholette}The Dagum and Cholette benchmarking method \cite{DagumCholette(94)}
uses
 the following three stochastic equations:
\begin{align}
\label{DC1}&Y^{H}_{O}=Hb+Z\delta+\theta+\epsilon_{H},
\hspace{2mm}\mathbb{E}(\epsilon_{H})=0,\hspace{2mm}\mathbb{E}(\epsilon_{H}
\epsilon_{H}^{'})=V_{\epsilon_{H}},\\
\label{DC2}&Y^{L}_{O}=JZ\delta+J\theta+\epsilon_{L},
\hspace{2mm}\mathbb{E}(\epsilon_{L})=0,\hspace{2mm}\mathbb{E}(\epsilon_{L}
\epsilon_{L}^{'})=V_{\epsilon_{L}},\\
\label{DC3}&S\theta=\eta,
\hspace{2mm}\mathbb{E}(\eta)=0,\hspace{2mm}\mathbb{E}(\eta\eta^{'})=V_{\eta}
\end{align}
The above equations are now discussed in the setting of quarterly to annual GDP benchmarking.

Equation \ref{DC1} decomposes the observed quarterly process into its true unobserved quarterly process $(Y^{Q}_{T}=Z\delta+\theta)$ and deterministic  $(Hb)$ and stochastic $(\epsilon_{H})$ disturbance terms. Typically $H$ is a vector of ones and $b$ a constant column vector forming a bias term capturing the average difference between the
observed quarterly $(Y^{Q}_{O})$ and annual $(Y^{A}_{O})$ series. $Z$ is an $n\times p$ matrix of known regressors and $\delta$ a $p\times 1$ vector of unknown coefficients modelling calendar effects. $\theta$ typically has an ARIMA structure; this is discussed below.

Equation \ref{DC2} decomposes the observed annual series $(Y^{A}_{O})$ into its true unobserved annual series
$(Y^{A}_{T}=JZ\delta+J\theta)$ and a disturbance term $(\epsilon_{L})$. $J$ is an annualising matrix equivalent to matrix $B{'}$ from the Denton method. The disturbance component $\epsilon_{L}$ is assumed to be Gaussian noise.

Matrix $S$ in equation \ref{DC3} transforms the stochastic component $\theta$ into a stationary time series. Set $\theta_{t}=\upsilon_{t}+\gamma_{t}+\epsilon_{t}$, with $\upsilon_{t}$ being approximately linear, i.e. $\upsilon_{t}\approx a+bt$,
$\gamma_{t}$ capturing quarterly seasonality and $\epsilon_{t}$ being Gaussian
random noise. In this scenario, to make $\theta$ stationary, the following matrix is required:
\begin{align}
S=
\left(\begin{array}{ccccccccc}
1&\hspace{-2.5mm}-1&0&0&\hspace{-2.5mm}-1&1&0&0&\ldots \\
0&1&\hspace{-2.5mm}-1&0&0&\hspace{-2.5mm}-1&1&0&\ldots \\
0&0&1&\hspace{-2.5mm}-1&0&0&\hspace{-2.5mm}-1&1&\ldots \\
\vdots&\vdots&\vdots&\vdots&\vdots&\vdots&\vdots&\vdots&\ddots
\end{array}\right)_{n-5\times n}
\end{align}
This is equivalent to applying the differencing operators $(1-L)$ and $(1-L^{4})$ to $\theta$. They remove linear and seasonal components from the series respectively, with $L$ denoting the lag operator, i.e. $L\theta_{t}=\theta_{t-1}$.

Model \ref{DC1}-\ref{DC3} can be written more concisely as:
\begin{align}
\label{DC4}
\left(\begin{array}{c}
Y^{H}_{O} \\
Y^{L}_{O} \\
0 \\
\end{array}\right)
=
\left(\begin{array}{ccc}
H&Z&I_{n} \\
0&JZ&J \\
0&0&S \\
\end{array}\right)
\left(\begin{array}{c}
b \\
\delta \\
\theta \\
\end{array}\right)
+
\left(\begin{array}{c}
\epsilon_{H} \\
\epsilon_{L} \\
-\eta \\
\end{array}\right),
\end{align}
or equivalently
\begin{align}
\label{DC5}
y=X\alpha+e,
\hspace{2mm}\mathbb{E}(e)=0,\hspace{2mm}V_{e}:=\mathbb{E}(ee^{'}
)=block(V_{\epsilon_{H}},V_{\epsilon_{L}},V_{\eta}),
\end{align}
$block(.,.,.)$ denotes a block diagonal matrix. Dagum and Cholette \cite{DagumCholette(94)} provided the following solution:
\begin{align}
\label{DC6}
\hat{\alpha}=(X^{'}V_{e}^{-1}X)^{-1}X^{'}V_{e}^{-1}y,
\end{align}

The benchmarked estimate is given by $\hat{\beta}=X^{*}\hat{\alpha}$, where $X^{*}=[0\hspace{1mm}Z\hspace{1mm}I_{n}]$. The following expresses the solution in a form consistent with  equation \ref{benchmarkingsolutiongeneral}:
\begin{align}
\label{DagumCholettesolutiongernalform}
\hat{Y}^{H}_{T}
=
\left(\begin{array}{ccc}
O & Z &I_{n} \\
\end{array}\right)
\left(\begin{array}{c}
\hat{b} \\
\hat{\delta} \\
\hat{\theta} \\
\end{array}\right)
=
\left(\begin{array}{ccc}
O & Z &I_{n} \\
\end{array}\right)
(X^{'}V^{-1}_{\mu}X)^{-1}X^{'}V^{-1}_{\mu}
\left(\begin{array}{c}
Y^{H}_{O} \\
Y^{L}_{O} \\
0 \\
\end{array}\right)
\end{align}
In Dagum and Cholette benchmarking the matrices $S, V_{\epsilon_{H}}, V_{\epsilon_{L}}$ and $V_{\eta}$ need to be estimated. To circumvent these estimation difficulties NSIs usually simplify the above model. The behaviour describing the unobserved stochastic component $\theta$ is ignored, i.e equation \ref{DC3} is removed. Since NSIs usually implement binding benchmarking $\epsilon_{L}=0$. Finally $\epsilon_{H}$ is modelled as an $AR(1)$ process. For practical implementation of Dagum and Cholette benchmarking, \cite*[Chapter~3]{DagumCholetteBook(2006)} recommend setting the $AR(1)$ parameter value between $0.7$ and $0.9$ for monthly series and between $0.7^3$ and $0.9^3$ for quarterly series. For monthly and quarterly time series, the ONS uses parameter values of $0.9$ and $0.9^3$ respectively \cite{BrownPS2012}. Naturally such adjustments can in some cases have a negative impact on the accuracy of the benchmarking process.

\subsection{Wavelets}\label{WaveletsSection}
Stationarity underpins many time series methods; this assumption is often unreasonable. Wavelets's time/frequency localisation enable segmentation of data over various frequency/time levels thus providing a framework to jointly analyse high (i.e quarterly data) and low (i.e. annual data) frequency series. While wavelets have facilitated recent advances in time series, i.e. alternative modelling of non stationary processes \cite{Nason(99)}, their primary use lies in non parametric regression and involves removing noise from a statistical process in a non parametric setting \cite{Donoho(94)}. Subsequent sections show the combination of a strict benchmarking and thresholding (denoising) step produces a benchmarking procedure which can outperform those currently used.

\subsubsection{Unbalanced Haar Wavelets}\label{UBHW}
The remainder of this section discusses Unbalanced Haar (UH) wavelets \cite{Fryzlewicz(2007)}. Data sets observed are typically non dyadic in length (i.e. $n\neq 2^{J}, J\in\mathbb{N}$). UH wavelets \cite{Fryzlewicz(2007)} are a generalisation of Haar wavelets \cite{Daubechies(92)} and enable the transformation of such non dyadic data sets into the wavelet domain. While discontinuities in Haar basis functions occur in the middle of their support (see Figure \ref{haarwavelets}), UH basis functions have discontinuities at arbitrary locations (see Figure \ref{unbalancedhharwaveletssegmentation}). Consequently high and low frequency data sets with arbitrary lengths/factor differences can be jointly analysed.

Consider the set $\left\lbrace 1,\ldots,n \right\rbrace$. The elementary father wavelet $\varphi^{-1,1}(t)$ is defined as:
\begin{align}
\varphi^{-1,1}(t)=\frac{1}{\sqrt{n}}1[1\leq t \leq n]
\end{align}
Let $s^{j,k} < b^{j,k} < e^{j,k}$ denote the startpoint, breakpoint and endpoint of a mother wavelet at scale level $j$ and translation level $k$. The mother wavelet $\varphi_{s^{j,k},b^{j,k},e^{j,k}}(t)$ is defined as follows (see Figure \ref{unbalancedhharwaveletssegmentation}):
\begin{align}
\varphi_{s^{j,k},b^{j,k},e^{j,k}}(t)&=\left[\frac{1}{b^{j,k}-s^{j,k}+1}-\frac{1}
{e^{j,k}-s^{j,k}+1}\right]^{\frac{1}{2}}1\left[s^{j,k} \leq t \leq b^{j,k}
\right]  \\
&- \left[
\frac{1}{e^{j,k}-b^{j,k}}-\frac{1}{e^{j,k}-s^{j,k}+1}\right]^{\frac{1}{2}}1\left
[b^{j,k}+1 \leq t \leq e^{j,k} \right] \nonumber
\end{align}
Given $\varphi^{j,k}(t):=\varphi_{s^{j,k},b^{j,k},e^{j,k}}(t)$, its two daughter wavelets $\varphi^{j+1,2k-1}(t)$, $\varphi^{j+1,2k}(t)$ (mother wavelets existing on higher frequency levels) with arbitrary breakpoints $b^{j+1,2k-1}$ (where $ s^{j,k}<b^{j+1,2k-1}<b^{j,k})$, $b^{j+1,2k}$ (where $ b^{j,k}<b^{
j+1,2k}<e^{j,k})$ are obtained as follows:
\begin{align}
&\varphi^{j+1,2k-1}(t)=\varphi_{s^{j,k},b^{j+1,2k-1},b^{j,k}}(t) \\
&\varphi^{j+1,2k}(t)=\varphi_{b^{j,k},b^{j+1,2k},e^{j,k}}(t)
\end{align}
This recursive process continues until an orthonormal wavelet basis is formed. Selecting appropriate breakpoints $\left\lbrace b^{0,1}, b^{1,1}, b^{1,2},\ldots \right\rbrace$ to ensure benchmarking can be performed is discussed in subsequent sections.

For a given set of startpoints $\left\lbrace s^{0,1}, s^{1,1}, s^{1,2},\ldots\right\rbrace$, breakpoints $\left\lbrace b^{0,1}, b^{1,1}, b^{1,2},\ldots\right\rbrace$ and endpoints $\left\lbrace e^{0,1}, e^{1,1}, e^{1,2},\ldots\right\rbrace$ the discrete unbalanced Haar transform (DUHT) of a series $Y=\left\lbrace Y_{t} \right\rbrace_{t=1}^{n}$ is defined as:
\begin{align}
w(j,k):=<Y,\varphi^{j,k}>=\sum\limits_{t=1}^{n}Y_{t}\varphi^{j,k}(t), \hspace{2mm}j=-1,\ldots,J \hspace{2mm}\text{and }k=1,\ldots k_{j}
\end{align}
To shorten notation $w(j,k)$'s and $\varphi^{j,k}$'s dependence on $\left\lbrace s^{j,k},b^{j,k},e^{j,k} \right\rbrace$ is implicit. In particular $w(-1,1)=\frac{1}{\sqrt{n}}\sum\limits_{t=1}^{n}Y_{t}$ denotes the elementary father wavelet coefficient; it summarises the average behaviour of the time series. Mother wavelet coefficients $w(j,k), j\geq 0$ provide information describing local features. Larger values of $j$ make the region of the time series considered narrower while $k$ determines the position considered on the time scale. The following resynthesises the original series $Y$ from the set of wavelet coefficients.
\begin{align}
\label{waveletrepresentation}
Y_{t}=w(-1,1)\varphi^{j,k}(t)+\sum\limits_{j=-1}^{J}\sum\limits_{k=1}^{k_{j}}w(j,k)\varphi^{j,k}(t), \hspace{2mm} t=1,\ldots,n
\end{align}
Equation \ref{waveletrepresentation} expresses $Y$ as a weighted linear combination of the elementary father wavelet and mother wavelets across various frequency and translation levels. Weights are given by their corresponding wavelet coefficients. This allows reconstruction of the benchmarked series after wavelet analysis.

\newgeometry{left=0.5cm,bottom=0.5cm,top=0.5cm}
\pgfplotsset{width=6cm,compat=newest}
\begin{figure}[H]
\centering
\begin{tikzpicture}[scale=0.7]
\begin{axis}[title={$\varphi^{-1,1}(.)$},
scale only axis,
ymin=-1.2,
ymax=1.2,
xmin=0,
xmax=512,
xtick={0,512}
]
\addplot[color=red,line width=3pt] coordinates {
(0, 0)
(0, 1)
(512,1)
(512,0)
};
\draw ({rel axis cs:0,0} |- {axis cs:0,0}) -- ({rel axis cs:1,0} |- {axis
cs:0,0});
\end{axis}
\begin{scope}[xshift=7cm]
\begin{axis}[title={$\varphi^{0,1}(.)$},
scale only axis,
ymin=-1.2,
ymax=1.2,
xmin=0,
xmax=512,
xtick={0,256,600}
]
\addplot[color=red,line width=3pt] coordinates {
(0, 0)
(0, 1)
(256,1)
(256,-1)
(512,-1)
(512,0)
};
\draw ({rel axis cs:0,0} |- {axis cs:0,0}) -- ({rel axis cs:1,0} |- {axis
cs:0,0});
\end{axis}
\end{scope}
\begin{scope}[xshift=14cm]
\begin{axis}[title={$\varphi^{1,1}(.)$},
scale only axis,
ymin=-1.2,
ymax=1.2,
xmin=0,
xmax=512,
xtick={128,256,512}
]
\addplot[color=red,line width=3pt] coordinates {
(0,0)
(0, 1)
(128,1)
(128,-1)
(256,-1)
(256,0)
(512,0)
};
\draw ({rel axis cs:0,0} |- {axis cs:0,0}) -- ({rel axis cs:1,0} |- {axis
cs:0,0});
\end{axis}
\end{scope}
\begin{scope}[xshift=21cm]
\begin{axis}[title={$\varphi^{1,2}(.)$},
scale only axis,
ymin=-1.2,
ymax=1.2,
xmin=0,
xmax=512,
xtick={0,256,384,512}
]
\addplot[color=red,line width=3pt] coordinates {
(0,0)
(256,0)
(256,1)
(384,1)
(384,-1)
(512,-1)
(512,0)
};
\draw ({rel axis cs:0,0} |- {axis cs:0,0}) -- ({rel axis cs:1,0} |- {axis
cs:0,0});
\end{axis}
\end{scope}
\end{tikzpicture}
\begin{tikzpicture}[scale=0.7]
\begin{axis}[title={$\varphi^{2,1}(.)$},
scale only axis,
ymin=-1.2,
ymax=1.2,
xmin=0,
xmax=512,
xtick={0,64,128,512}
]
\addplot[color=red,line width=3pt] coordinates {
(0, 0)
(0, 1)
(64,1)
(64,-1)
(128,-1)
(128,0)
(512,0)
};
\draw ({rel axis cs:0,0} |- {axis cs:0,0}) -- ({rel axis cs:1,0} |- {axis
cs:0,0});
\end{axis}
\begin{scope}[xshift=7cm]
\begin{axis}[title={$\varphi^{2,2}(.)$},
scale only axis,
ymin=-1.2,
ymax=1.2,
xmin=0,
xmax=512,
xtick={0,128,192,256,512}
]
\addplot[color=red,line width=3pt] coordinates {
(0,0)
(128, 0)
(128,1)
(192,1)
(192,-1)
(256,-1)
(256,0)
(512,0)
};
\draw ({rel axis cs:0,0} |- {axis cs:0,0}) -- ({rel axis cs:1,0} |- {axis
cs:0,0});
\end{axis}
\end{scope}
\begin{scope}[xshift=14cm]
\begin{axis}[title={$\varphi^{2,3}(,)$},
scale only axis,
ymin=-1.2,
ymax=1.2,
xmin=0,
xmax=512,
xtick={0,256,320,384}
]
\addplot[color=red,line width=3pt] coordinates {
(0, 0)
(256, 0)
(256,1)
(320,1)
(320,-1)
(384,-1)
(384,0)
(512,0)
};
\draw ({rel axis cs:0,0} |- {axis cs:0,0}) -- ({rel axis cs:1,0} |- {axis
cs:0,0});
\end{axis}
\end{scope}
\begin{scope}[xshift=21cm]
\begin{axis}[title={$\varphi^{2,4}(.)$},
scale only axis,
ymin=-1.2,
ymax=1.2,
xmin=0,
xmax=512,
xtick={0,384,448,512}
]
\addplot[color=red,line width=3pt] coordinates {
(0, 0)
(384, 0)
(384,1)
(448,1)
(448,-1)
(512,-1)
(512,0)
};
\draw ({rel axis cs:0,0} |- {axis cs:0,0}) -- ({rel axis cs:1,0} |- {axis
cs:0,0});
\end{axis}
\end{scope}
\end{tikzpicture}
\caption{Haar wavelet segmentation for a time series of length $t=512$ for frequency scales $-1,0,1,2$. For illustration purposes these wavelets have been rescaled.}
\label{haarwavelets}
\begin{tikzpicture}[scale=0.7]
\begin{axis}[title={$\varphi^{-1,1}(.)$},
scale only axis,
ymin=-1.2,
ymax=1.2,
xmin=0,
xmax=600,
]
\addplot[color=blue,line width=3pt] coordinates {
(0, 0)
(0, 1)
(600,1)
(600,0)
};
\draw ({rel axis cs:0,0} |- {axis cs:0,0}) -- ({rel axis cs:1,0} |- {axis
cs:0,0});
\end{axis}
\begin{scope}[xshift=7cm]
\begin{axis}[title={$\varphi^{0,1}(.)$},
scale only axis,
ymin=-1.2,
ymax=1.2,
xmin=0,
xmax=600,
xtick={0,88,600}
]
\addplot[color=blue,line width=3pt] coordinates {
(0, 0)
(0, 1)
(88,1)
(88,-1)
(600,-1)
(600,0)
};
\draw ({rel axis cs:0,0} |- {axis cs:0,0}) -- ({rel axis cs:1,0} |- {axis
cs:0,0});
\end{axis}
\end{scope}
\begin{scope}[xshift=14cm]
\begin{axis}[title={$\varphi^{1,1}(.)$},
scale only axis,
ymin=-1.2,
ymax=1.2,
xmin=0,
xmax=600,
xtick={24,88,600}
]
\addplot[color=blue,line width=3pt] coordinates {
(0,0)
(0, 1)
(24,1)
(24,-1)
(88,-1)
(88,0)
(600,0)
};
\draw ({rel axis cs:0,0} |- {axis cs:0,0}) -- ({rel axis cs:1,0} |- {axis
cs:0,0});
\end{axis}
\end{scope}
\begin{scope}[xshift=21cm]
\begin{axis}[title={$\varphi^{1,2}(.)$},
scale only axis,
ymin=-1.2,
ymax=1.2,
xmin=0,
xmax=600,
xtick={0,88,344,600}
]
\addplot[color=blue,line width=3pt] coordinates {
(0,0)
(88,0)
(88,1)
(344,1)
(344,-1)
(600,-1)
(600,0)
};
\draw ({rel axis cs:0,0} |- {axis cs:0,0}) -- ({rel axis cs:1,0} |- {axis
cs:0,0});
\end{axis}
\end{scope}
\end{tikzpicture}
\begin{tikzpicture}[scale=0.7]
\begin{axis}[title={$\varphi^{2,1}(.)$},
scale only axis,
ymin=-1.2,
ymax=1.2,
xmin=0,
xmax=88,
xtick={0,8,24,88}
]
\addplot[color=blue,line width=3pt] coordinates {
(0, 0)
(0, 1)
(8,1)
(8,-1)
(24,-1)
(24,0)
(88,0)
};
\draw ({rel axis cs:0,0} |- {axis cs:0,0}) -- ({rel axis cs:1,0} |- {axis
cs:0,0});
\end{axis}
\begin{scope}[xshift=7cm]
\begin{axis}[title={$\varphi^{2,2}(.)$},
scale only axis,
ymin=-1.2,
ymax=1.2,
xmin=0,
xmax=88,
xtick={0,24,56,88}
]
\addplot[color=blue,line width=3pt] coordinates {
(0,0)
(24, 0)
(24,1)
(56,1)
(56,-1)
(88,-1)
(88,0)
};
\draw ({rel axis cs:0,0} |- {axis cs:0,0}) -- ({rel axis cs:1,0} |- {axis
cs:0,0});
\end{axis}
\end{scope}
\begin{scope}[xshift=14cm]
\begin{axis}[title={$\varphi^{2,3}(.)$},
scale only axis,
ymin=-1.2,
ymax=1.2,
xmin=88,
xmax=600,
xtick={88,216,344,600}
]
\addplot[color=blue,line width=3pt] coordinates {
(0, 0)
(88, 0)
(88,1)
(216,1)
(216,-1)
(344,-1)
(344,0)
(600,0)
};
\draw ({rel axis cs:0,0} |- {axis cs:0,0}) -- ({rel axis cs:1,0} |- {axis
cs:0,0});
\end{axis}
\end{scope}
\begin{scope}[xshift=21cm]
\begin{axis}[title={$\varphi^{2,4}(.)$},
scale only axis,
ymin=-1.2,
ymax=1.2,
xmin=88,
xmax=600,
xtick={88,344,472,600}
]
\addplot[color=blue,line width=3pt] coordinates {
(0, 0)
(344, 0)
(344,1)
(472,1)
(472,-1)
(600,-1)
(600,0)
};
\draw ({rel axis cs:0,0} |- {axis cs:0,0}) -- ({rel axis cs:1,0} |- {axis
cs:0,0});
\end{axis}
\end{scope}
\end{tikzpicture}
\caption{An example of an unbalanced Haar wavelet segmentation for a time series of length $t=600$ for frequency scales $-1,0,1,2$. For illustration purposes these wavelets have been rescaled.}
\label{unbalancedhharwaveletssegmentation}
\end{figure}

\newgeometry{left=2.5cm,right=2.5cm,bottom=2cm,top=2cm}
\begin{table}[H]
\begin{center}
\begin{tabular}{|c|c|c|c|c|c|c|c|c|c|c|}
\hline
UBHW length & $n_{0,1}$ & $n_{1,1}$ & $n_{1,1}^{+}$ & $n_{1,1}^{-}$ & $n_{2,1}$
& $n_{2,1}^{+}$ & $n_{2,1}^{-}$ & $n_{2,2}$ & $n_{2,2}^{+}$ & $n_{2,2}^{-}$  \\
\hline
Value & 600 & 600 & 88 & 512 & 88 & 24 &  64 & 512 & 256 & 256  \\
\hline
\end{tabular}

\begin{tabular}{|c|c|c|c|c|c|c|c|c|c|}
\hline
UBHW length & $n_{3,1}$ & $n_{3,1}^{+}$ & $n_{3,1}^{-}$ & $n_{3,2}$ &
$n_{3,2}^{+}$ & $n_{3,2}^{-}$ & $n_{3,3}=n_{3,4}$ & $n_{3,3}^{+}=n_{3,4}^{+}$ &
$n_{3,3}^{-}=n_{3,4}^{-}$  \\
\hline
Value & 24 & 8 & 16 & 64 & 32 & 32 & 256 & 128 & 128 \\
\hline
\end{tabular}
\end{center}
\caption{Support of vectors used to perform the wavelet transform for non dyadic
data of length $600$}
\label{tablenondyadicsegmentation}
\end{table}

\section{Methodology}\label{Methodology}
In this section we discuss the selection of wavelet bases used to facilitate benchmarking. Elementary wavelet benchmarking is introduced along with an application to simulated data. Finally the additional issue of thresholding and its integration with seasonal adjustment is considered.

\subsection{Wavelet Basis Selection}

\subsubsection{Forming a Basis for Non Dyadic Data using Unbalanced Haar Wavelets}\label{formingbasisnondyadicdata}
NSIs regularly revise published time series and since published economic data impacts decisions implemented by policy makers, producing a stable benchmarked series is important. To reduce benchmarked series sensitivity to such adjustments, observed time series could be transformed into the wavelet domain using a segmentation that spreads latter regions of the time series across as many frequency levels as possible. The formation of such a basis is outlined as follows. At each iteration the positive region of the mother wavelet being considered is segmented into a daughter wavelet with the largest possible dyadic region and non dyadic positive region. Its negative region is segmented into a daughter wavelet with positive and negative regions of equal length (Haar segmentation).

More formally consider the support of $\varphi^{j,k}$ along with the support of its positive and negative regions. Denote their cardinality by $n_{j,k}$, $n^{+}_{j,k}$ and $n^{-}_{j,k}$ respectively. For the father wavelet $\varphi^{-1,1}$, $|supp(\varphi^{-1,1})|=n_{-1,1}=n^{+}_{-1,1}$ $(supp(f):=\left\lbrace x;f(x)\neq 0 \right\rbrace)$, with $n_{-1,1}$ being the length of the time series. $\varphi^{-1,1}$ is decomposed forming the mother wavelet $\varphi^{0,1}$, with $|supp(\varphi^{0,1})|=n_{0,1}=n_{-1,1}$. Setting $n^{-}_{0,1}=2^{\lfloor \text{log}_{2}(n_{0,1})\rfloor}$ ($\lfloor . \rfloor$ denotes the greatest integer function) ensures the negative region of $\varphi^{0,1}$ has the largest possible dyadic support. Consequently $n^{+}_{0,1}=n_{0,1}-n^{-}_{0,1}$; typically $n^{+}_{0,1}$ is non dyadic in length. Its corresponding region is segmented in a similar manner to $\varphi^{0,1}$ while regions of dyadic support (regions of $\varphi^{0,1}$ corresponding to $n^{-}_{0,1}$) are segmented using the Haar transform. This iterative process continues until a basis is formed.

Figure \ref{unbalancedhharwaveletssegmentation} illustrates an example of the above segmentation using UH wavelets with the frequency levels $-1,0,1,2$ considered. Table \ref{tablenondyadicsegmentation} records the support of these wavelets. To provide a comparison Figure \ref{haarwavelets} illustrates the Haar segmentation on the same frequency levels.

\subsubsection{Creating a Benchmarking Basis}\label{creatingbenchmarkingbasis}
The set of breakpoints $\left\lbrace b^{0,1}, b^{1,1}, b^{1,2},\ldots \right\rbrace$  determines the UH wavelet basis. Benchmarking requires the bases for low and high frequency processes to be comparable.

The low $\left(\left\lbrace Y^{L}_{t}\right\rbrace_{t=1}^{m}\right)$ and high $\left(\left\lbrace Y^{H}_{t}\right\rbrace_{t=1}^{n}\right)$ frequency series are observed, with $n=km$ and $k$ being the factor difference. Let
$L_{BP}=\left\lbrace b^{0,1}_{L}, b^{1,1}_{L}, b^{1,2}_{L}, \ldots, b^{J_{L},k_{J_{L}}}_{L}\right\rbrace$ and $H_{BP}=\left\lbrace b^{0,1}_{H},\ldots, b^{J_{L},k_{J_{L}}}_{H}, b^{J_{L}+1,1}_{H},\ldots b^{J_{H},k_{J_{H}}}_{H}\right\rbrace$ represent the low and high frequency series set of breakpoints respectively.

The set of breakpoints $L_{BP}$ is selected by the method described in section \ref{formingbasisnondyadicdata}. Breakpoints for $H_{BP}$ with overlapping frequency levels with $L_{BP}$ are defined as:
\begin{align}
b^{j,k}_{H}=kb^{j,k}_{L}, \hspace{2mm} j=1,\ldots,J_{L}, \hspace{2mm}k=1,\ldots,k_{j}
\end{align}
Remaining breakpoints $\left\lbrace b^{J_{L}+1,1}_{H},\ldots b^{J_{H},k_{J_{H}}}_{H}\right\rbrace$ can be chosen arbitrarily as they exist on frequency levels not affected by elementary benchmarking. To maintain consistency the procedure in Section \ref{formingbasisnondyadicdata} is used. The sets $L_{BP}$, $H_{BP}$ provide the foundation required to perform elementary wavelet benchmarking.

\subsection{Elementary Wavelet Benchmarking}\label{waveletbenchmarking}
Consider quarterly to annual GDP binding benchmarking. The quarterly $\left\lbrace Y^{Q}_{O,t} \right\rbrace_{t=1}^{n}$ and annual $\left\lbrace Y^{A}_{O,t}\right\rbrace_{t=1}^{m}$ GDP series are observed ($m=\frac{n}{4}$). Both series are expressed in the form described by equation \ref{waveletrepresentation}:
\begin{align*}
&Y^{Q}_{O,t}=\tilde{w}^{Q}(-1,0)\varphi_{-1,0}^{Q}(t)+\sum\limits_{j=1}^{J+2}\sum\limits_{k=0}^{k_{j}}\tilde{w}^{Q}(j,k)\varphi^{Q}_{j,k}(t), \hspace{2mm}t=1,\ldots,n \\
&Y^{A}_{O,t}=w^{A}(-1,0)\varphi_{-1,0}^{A}(t)+\sum\limits_{j=1}^{J}\sum\limits_{k=0}^{k_{j}}w^{A}(j,k)\varphi^{A}_{j,k}(t), \hspace{2mm} t=1,\ldots,m
\end{align*}
The construction of wavelet functions $\varphi^{Q}_{j,k}(.)$ and $\varphi^{A}_{j,k}(.)$ defined on the sets $\left\lbrace 1,\ldots,n \right\rbrace$ and $\left\lbrace 1,\ldots,m \right\rbrace$ respectively is discussed in Section \ref{UBHW}. $\tilde{w}^{Q}(-1,0)$, $\tilde{w}^{Q}(j,k)$ and $w^{A}(-1,0)$, $w^{A}(j,k)$ denote noisy and non noisy wavelet coefficients from the quarterly and annual time series respectively. $J$ is the highest frequency level of the annual time series; the quarterly time series has $2$ additional frequency levels.

Quarterly wavelet coefficients existing on lower frequency levels of the wavelet domain have corresponding annual wavelet coefficients with similar interpretations i.e. coefficients existing on frequency levels $j=-1,\ldots,J$. A comparison of elementary quarterly and annual father wavelet coefficients illustrates this:
\begin{align}
\label{noisyquarterlycoefficient}
\tilde{w}^{Q}(-1,1)=&\frac{1}{\sqrt{n}}\sum\limits_{t=1}^{n}Y^Q_{O,t}, \\
w^{A}(-1,1)=&\frac{1}{\sqrt{m}}\sum\limits_{t=1}^{m}Y^{A}_{O,t}=\frac{1}{\sqrt{m}}\sum\limits_{t=1}^{m}Y^{A}_{T,t}, \hspace{2mm} \text{since the annual GDP series is non noisy} \nonumber \\
\label{nonnoisyannualcoefficient}
=&\frac{1}{\sqrt{m}}\sum\limits_{t=1}^{m}\sum\limits_{j=4t-3}^{4t}Y^{Q}_{T,j}=\frac{1}{\sqrt{m}}\sum\limits_{t=1}^{n}Y^{Q}_{T,t}=\frac{2}{\sqrt{n}}\sum\limits_{t=1}^{n}Y^{Q}_{T,t}=2w^{Q}(-1,1), \hspace{2mm} \text{since } n=4m.
\end{align}
This illustrates the key idea of elementary wavelet benchmarking; replacing $\tilde{w}^{Q}(j,k)$ with $\frac{1}{2}w^{A}(j,k)$ for wavelet coefficients on frequency levels $j=-1,\ldots,J$ produces the benchmarked series  $\left\lbrace \hat{Y}^{Q}_{T,t}\right\rbrace_{t=1}^{n}$:
\begin{align}
\label{benchmarkedseries}
\hat{Y}^{Q}_{T,t}=\underbrace{\frac{w^{A}(-1,0)}{2}\varphi_{-1,0}^{Q}(t)+\sum\limits_{j=0}^{J}\sum\limits_{k=0}^{k_{j}}\frac{w^{A}(j,k)}{2}\varphi^{Q}_{j,k}(t)}_{Y^{A,Q}_{T,t}}+\underbrace{\sum\limits_{j=J+1}^{J+2}\sum\limits_{k=0}^{k_{j}}\tilde{w}^{Q}(j,k)\varphi_{j,k}(t)}_{R^{J+1,J+2}_{Y^{Q}_{O,t}}}
\end{align}
Equation \ref{benchmarkedseries} decomposes the benchmarked series into two components $Y^{A,Q}_{T,t}$ and $R^{J+1,J+2}_{Y^{Q}_{O,t}}$. $Y^{A,Q}_{T,t}$ expresses the non noisy annual series on a quarterly time scale with no intra-annual fluctuations (i.e. quarterly values in a given year take the same value). $R^{J+1,J+2}_{Y^{Q}_{O,t}}$ isolates fluctuations unique to the quarterly time series. Since it exists on the frequency levels $J+1$ and $J+2$ it has no impact on the annualised version of $\hat{Y}^{Q}_{T,t}$. Therefore the benchmark constraint is satisfied, i.e:
\begin{align*}
\hat{Y}^{A}_{T,t}=\sum\limits_{j=4t-3}^{4t}\hat{Y}^{Q}_{T,t}=Y^{A}_{T,t}
\end{align*}
Thresholding  $\left\lbrace R^{J+1,J+2}_{Y^{Q}_{O,t}} \right\rbrace_{t=1}^{n}$ can further improve the estimation of $\left\lbrace \hat{Y}^{Q}_{T} \right\rbrace_{t=1}^{n}$; as discussed in later sections.

Elementary wavelet benchmarking is expressed in a form consistent with equation \ref{benchmarkingsolutiongeneral} as follows. While computationally inefficient, the wavelet transform can be expressed as an orthogonal matrix; see \cite*[Chapter~2]{Nason(2008)} for details. Suppose $W^{Q}$ and $W^{A}$ transform the quarterly and annual series into the wavelet domain respectively:
\begin{align}
\tilde{y}^{Q}=W^{Q}Y^{Q}_{O}, \hspace{2mm} y^{A}=W^{A}Y^{A}_{O}
\end{align}
Decomposing $W^{Q}$ into low $(W^{Q,A})$ and high $(W^{Q,Q})$ frequency components enable the low $(\tilde{y}^{Q,A})$ and high $(\tilde{y}^{Q,Q})$ frequency wavelet coefficients to be obtained:
\begin{align*}
\tilde{y}^{Q}=\left[
\begin{array}{c}
\tilde{y}^{Q,A} \\
\tilde{y}^{Q,Q}
\end{array}\right]
=
\left[
\begin{array}{c}
W^{Q,A} \\
W^{Q,Q}
\end{array}
\right]
Y^{Q}_{O}
\end{align*}
The non noisy annual wavelet coefficients $y^{A}$ can be incorporated into the noisy quarterly wavelet coefficients $\tilde{y}^{Q}$ as follows:
\begin{align*}
\hat{y}^{Q}=\left[
\begin{array}{c}
\frac{1}{c}y^{A} \\
\tilde{y}^{Q,Q}
\end{array}\right]
=\left[
\begin{array}{ccc}
0 & 0 & \frac{1}{c}I \\
0 & I & 0
\end{array}
\right]
\left[
\begin{array}{c}
\tilde{y}^{Q,A} \\
\tilde{y}^{Q,Q} \\
y^{A}
\end{array}
\right]
\end{align*}
As seen from equation \ref{nonnoisyannualcoefficient}, in the example of quarterly to annual benchmarking $c=2$. The benchmarked series $\left\lbrace \hat{Y}^{H}_{T,t} \right\rbrace_{t=1}^{n}$ is then calculated as follows:
\begin{align*}
\hat{Y}^{H}_{T}=\left(W^{Q}\right)^{T}\hat{y}^{Q}=\left(W^{Q}\right)^{T}\left[
\begin{array}{c}
\frac{1}{c}y^{A} \\
\tilde{y}^{Q,Q}
\end{array}\right]
=\underbrace{\left(W^{Q}\right)^{T}\left[
\begin{array}{ccc}
0 & 0 & \frac{1}{c}I \\
0 & I & 0
\end{array}\right]\left[
\begin{array}{cc}
W^{Q} & 0 \\
0 & W^{A}
\end{array}\right]}_{\text{Elementary Wavelet Benchmarking Matrix}}\left[
\begin{array}{c}
Y^{Q}_{O} \\
Y^{A}_{O}
\end{array}\right]
\end{align*}

\subsubsection{Example Application of Elementary Wavelet Benchmarking} \label{applicationelementarwaveletbenchmarking}
\begin{figure}[H]
\begin{center}
\includegraphics[width=\textwidth, height=15cm]{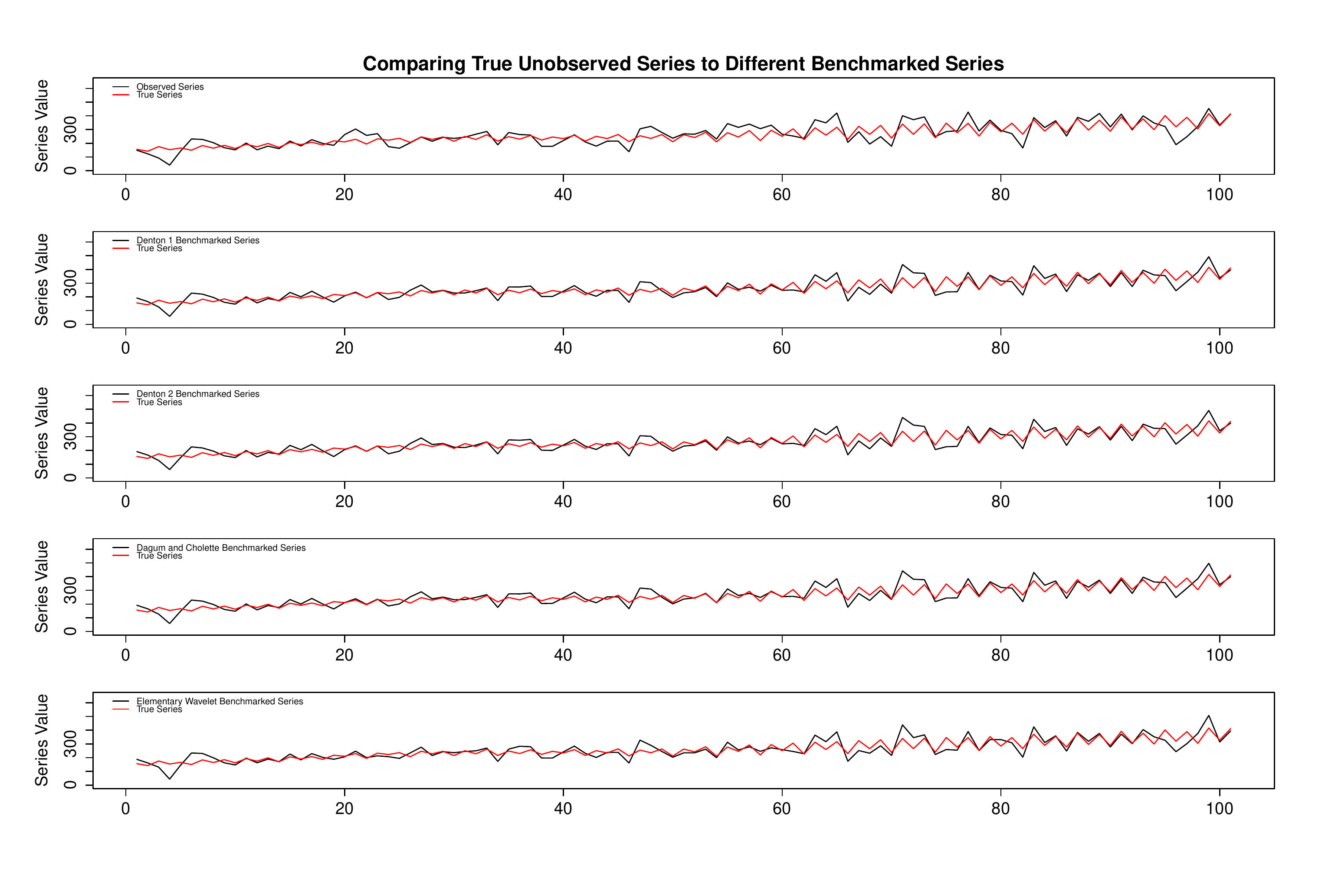}
\caption{Comparing the different benchmarking methods to the true unobserved series for a single simulation for dyadic quarterly and annual data.}
\label{SingleSimulationCompareBenchmarking}
\end{center}
\end{figure}

In certain circumstances performing elementary wavelet benchmarking is sufficient (i.e. small survey error). Elementary wavelet benchmarking is applied to simulated quarterly/annual GDP  time series as an example of this. For simplicity, data is dyadic allowing the Haar transform to be applied. Since structural time series (STS) models \cite*[Chapter~3]{DurbinKoopman(2001)} adequately describe many economic processes they are used to generate simulations and in particular do not conform to any of the chosen methodologies providing a valid comparison, not biased to any of the underlying benchmarking methods. 500 simulated data sets were generated by the STS model in Appendix \ref{appendixSTSmodel}. Parameter and initialisation values are reported in Table \ref{parametervalues1}. Average MSE values for the original and four different benchmarking methods are reported in Table \ref{elementary500simulationstable1}. Finally for a single simulated series Figure \ref{SingleSimulationCompareBenchmarking} provides an illustrated comparison between a subsection of the true unobserved series and each of the benchmarked series.
\begin{table}[H]
\begin{center}
\begin{tabular}{|c|c|}
\hline
Series Type & Average MSE Values \\
\hline
Original &  2419.84 \\
Denton 1 &  1208.75 \\
Denton 2 &  1252.84 \\
Dagum and Cholette & 1203.51 \\
Elementary Wavelet Benchmarking & 1253.77 \\
\hline
\end{tabular}
\caption{Average MSE values comparing benchmarking methods of 500 dyadic quarterly/annual series.}
\label{elementary500simulationstable1}
\end{center}
\end{table}
Table \ref{elementary500simulationstable1} indicates elementary wavelet benchmarking performs similarly to current methods. As will be illustrated in Section \ref{dataanalysissection} taking the presence of noise into account and thresholding affected wavelet coefficients produces a benchmarking method which can outperform both Denton and Dagum and Cholette benchmarking.

\subsection{Thresholding}\label{thresholding}
In the previous example the benchmarked series $\left\lbrace \hat{Y}^{Q}_{T,t} \right\rbrace_{t=1}^{n}$ was decomposed into a low frequency non noisy component $\left\lbrace Y^{A,Q}_{T,t} \right\rbrace_{t=1}^{n}$ and a high frequency noisy component $\left\lbrace R^{J+1,J+2}_{Y^{Q}_{O,t}}\right\rbrace_{t=1}^{n}$. Thresholding wavelet coefficients corresponding to $\left\lbrace R^{J+1,J+2}_{Y^{Q}_{O,t}}\right\rbrace_{t=1}^{n}$ produces a more reliable benchmarked series, as it removes spurious noise.

Technical details of thresholding are available in \cite*[Chapter 6]{Vidakovic(99)}; however two features of the error term are important.  Firstly its structure, in real data sets, random components typically exhibit some form of autocorrelation. Hence iid Gaussian noise is not appropriate. Thus simulations in this paper use an $ARMA(1,1)$ process to generate disturbance terms. Consequently, wavelet coefficients on a given frequency level are correlated; hence thresholding based on Stein's unbiased risk estimator (SURE)  \cite*[Chapter~10]{PercivalWalden(2000)} is used. Secondly, to estimate the error terms variance in the wavelet domain, we use the method in Percival \cite{Percival(95)} for estimating the variance across different frequency levels.

\subsubsection{Thresholding Framework}\label{thresholdingframework}
Suppose $Y_{O,t}=Y_{T,t}+\epsilon_{t}$, $t=1,\ldots,n$ is observed, with $\epsilon_{t}$ being an error term. Transforming $Y_{O}=\left[Y_{O,1},\ldots,Y_{O,n}\right]^{'}$ into the wavelet domain using the orthogonal matrix $W$; we have
\begin{align}
w=WY_{O},
\end{align}
$w=\left[w_{1},\ldots,w_{n}\right]^{'}$. Typically either hard or soft thresholding \cite{Donoho(94)} is used. In this paper, we use soft thresholding with estimates obtained as follows:
\begin{align}
\label{softthresholding}
\hat{w}_{i}&=sgn(w_{i})\left[|w_{i}|-\lambda\right]\mathds{1}\left(|w_{i}|\geq\lambda\right),\\
&=\left[1-\frac{\lambda}{|w_{i}|}\right]\mathds{1}\left(|w_{i}|\geq\lambda\right)w_{i}
\end{align}
 $\lambda$ denotes the threshold value (parameter depending on the noise level), and $\mathds{1}(\cdot)$ is the indicator function.

If the magnitude of an observed wavelet coefficient is greater than $\lambda$ it is shrunk in magnitude by $\lambda$. Otherwise it is set to zero. As mentioned above $\lambda$ is estimated based on SURE; such an estimator depends on both the series length and variance of the noise term. In particular Percival's estimator based on the maximal overlap discrete wavelet transform is used to estimate the variance of noisy wavelet coefficients; \cite{Percival(95)} discusses this in more detail. Consequently $\lambda$ is a data dependent parameter, i.e $\lambda=\lambda(Y_{O})$ Using equation \ref{softthresholding}, estimates of the true wavelet coefficients are obtained as follows:
\begin{align}
\hat{w}
=
\underbrace{\left(\begin{array}{cccc}
t_{1}&0&\ldots&0 \\
0&\ddots&\ldots&\ldots \\
0&\ldots&\ddots&\ldots\\
0&\ldots&\ldots&t_{n}\\
\end{array}\right)}_{T_{w}}
w
\end{align}
The diagonal elements of $T_{w}$, are $t_{i}=\left[1-\frac{\hat{\lambda}}{|w_{i}|}\right]\mathds{1}\left(|w_{i}|\geq\hat{\lambda}\right)$, with $\hat{\lambda}$ being a threshold estimate. An estimate of the unobserved true series $Y_{T}$ is now obtained as:
\begin{align}
\label{thresholdgeneralsolution}
\hat{Y}_{T}=W^{'}T_{w}WY_{O}
\end{align}

\subsection{Alternative Seasonal Model}\label{alternativeseasonalmodelbenchmarking}
As seen in Section \ref{dataanalysissection}, in many cases the noisy high frequency series requires seasonal adjusting \cite*[Chapter~3]{DurbinKoopman(2001)} prior to benchmarking/thresholding and is reintroduced afterwards. The seasonal component is unknown and hence must be estimated. Time series data being studied in this paper are represented in state space form \cite*[Chapter~3]{DurbinKoopman(2001)}. To estimate the seasonal component we apply the Kalman smoother \cite*[Chapter~4]{DurbinKoopman(2001)}. A stochastic seasonal model taking the following form is used:
\begin{align}\label{stochasticseasonalmodel}
\gamma_{t+1}=-\sum\limits_{j=1}^{k-1}\gamma_{t+1-j}+\omega_{t},
\hspace{2mm}\omega_{t}\sim N(0,\sigma^{2}_{\omega})
\end{align}
However, the zero sum constraint of the seasonal component is violated $\left(\text{i.e. }\sum\limits_{j=1}^{k}\gamma_{t+1-j}\neq 0\right)$. Consequently the benchmark constraint will no longer be satisfied once the seasonal estimate is reintroduced into the series. Therefore the following representation, which allows the seasonal process to vary stochastically while ensuring the zero sum constraint is satisfied is considered.
\begin{align}\label{alternativeseasonalmodel}
\left(\begin{array}{c}
\gamma_{1,t+1} \\
\vdots \\
\gamma_{k,t+1} \\
\end{array}\right)
=
\left(\begin{array}{c}
\gamma_{1,t} \\
\vdots \\
\gamma_{k,t} \\
\end{array}\right)
+
\left(\begin{array}{c}
\omega_{1,t} \\
\vdots \\
\omega_{k,t} \\
\end{array}\right),
\end{align}
or equivalently,
\begin{align}
\mathbf{\gamma_{t+1}}=\mathbf{\gamma_{t}}+\mathbf{\omega_{t}}
\end{align}
In equation \ref{alternativeseasonalmodel} any season $j$ within a given year $t+1$ takes the value $\gamma_{j,t+1}$ and is equal to its value from the previous year $\gamma_{j,t}$ plus a disturbance term. One way to ensure the seasonally adjusted series satisfies the benchmark constraint is to define an appropriate correlation structure $\left(Var(\omega_{t})=\sigma_{\omega}^{2}\left(\mathbf{I_{k}}-\frac{1}{k}\mathbf{I_{
k\times 1}I_{k \times 1}^{'}}\right)\right)$ between the components of $\mathbf{\omega_{t}}$. Therefore the sum of each year's $k$ seasons is constant, i.e. the following holds:
\begin{align}
\label{alternativeseasonalmodel1}
\sum\limits_{j=1}^{k}\gamma_{j,t+1}=\sum\limits_{j=1}^{k}\gamma_{j,t}
=\ldots=\sum\limits_{j=1}^{k}\gamma_{j,0}
\end{align}
Imposing the above correlation structure results in $\mathbb{E}(\mathbf{I_{k\times 1}}^{'}\mathbf{\omega_{t}})=0=Var(\mathbf{I_{k\times 1}}^{'}\mathbf{\omega_{t}})$. This, along with the initialisation condition $\sum\limits_{j=1}^{k}\gamma_{j,0}=0$, forces the benchmark constraint to hold.

To maintain consistency, other components from structural time series models (i.e. trend, slope and error components) are represented in a similar form to equation \ref{alternativeseasonalmodel}. Such models are known as periodic structural time series; more information is provided in \cite{Tripodis(2004)}

\subsection{Wavelet Benchmarking Algorithm}\label{elementarybenchmarkinthresholdingsummary}
The following summarises wavelet benchmarking:
\par\vspace{2mm}
\textbf{input:} A high and low frequency series denoted $Y^{H}_{O}$ (length n) and $Y^{L}_{O}$ (length m) respectively.
\newline \textbf{ouput:} A benchmarked series $\hat{Y}^{H}_{T}$
\par\vspace{2mm}
\textbf{if} Seasonality is present \textbf{then}
\par Seasonally adjust the high frequency series:
\begin{center}
$\check{Y}^{H}_{O}=Y^{H}_{O}-\hat{\gamma}$, where $\hat{\gamma}$ denotes the estimated seasonal component
\end{center}
\textbf{else} Do not perform seasonal adjustment:
\begin{center}
$\check{Y}^{H}_{O}=Y^{H}_{O}$
\end{center}
\textbf{Transform $\check{Y}^{H}_{O}$ and $Y^{L}_{O}$ from the time to wavelet domain (Section \ref{WaveletsSection}):}
\newline Represent the wavelet transform for $\check{Y}^{H}_{O}$ and $Y^{L}_{O}$ by the orthogonal matrices $W^{H}$ and $W^{L}$ respectively. This produces the following vector of wavelet coefficients:
\begin{align*}
\left[
\begin{array}{c}
\tilde{y}^{H} \\
y^{L}
\end{array}\right]
=
\left[
\begin{array}{cc}
W^{H} & 0 \\
0 & W^{L}
\end{array}\right]
\left[
\begin{array}{c}
\check{Y}^{H}_{O}\\
Y^{L}_{O}
\end{array}\right]
\end{align*}
\par\vspace{2mm}
\textbf{Apply Elementary Benchmarking and Thresholding (Section \ref{waveletbenchmarking} and Section \ref{thresholding}):}
\newline $\tilde{y}^{H}$ is decomposed into a noisy low frequency $(\tilde{y^{H,L}})$ and high frequency $(\tilde{y^{H,H}})$ component:
\begin{align*}
\left[
\begin{array}{c}
\tilde{y}^{H,L} \\
\tilde{y}^{H,H} \\
y^{L}
\end{array}\right]
=
\left[
\begin{array}{c}
\tilde{y}^{H} \\
y^{L}
\end{array}\right]
\end{align*}
Applying elementary benchmarking results in the following set of high frequency wavelet coefficients:
\begin{align*}
\left[
\begin{array}{c}
\frac{1}{c}y^{L} \\
\tilde{y}^{H,H}
\end{array}\right]
=
\left[
\begin{array}{ccc}
0 & 0 & \frac{1}{c}I \\
0 & I & 0
\end{array}\right]
\left[
\begin{array}{c}
\tilde{y}^{H,L} \\
\tilde{y}^{H,H} \\
y^{L}
\end{array}\right]
\end{align*}
c represents the constant taking the scale difference between the high and low frequency series into account. Thresholding is applied to coefficients existing on high frequency regions i.e the coefficients $\tilde{y^{H,H}}$:
\begin{align*}
\hat{y}^{h}=
\left[
\begin{array}{c}
\frac{1}{c}y^{L}\\
\hat{y}^{H,H}
\end{array}\right]
&=
\left[
\begin{array}{cc}
I_{m} & 0_{m,n-m,} \\
0_{n-m,m} & T_{n-m,n-m}(\tilde{y}^{H,H})
\end{array}\right]
\left[
\begin{array}{c}
\frac{1}{c}y^{L} \\
\tilde{y}^{H,H}
\end{array}\right] \\
\end{align*}
$T_{n-m,n-m}(\tilde{y}^{H,H})$ is a data dependent matrix performing the thresholding operation.
\par\vspace{2mm}
\textbf{Transform the estimated high frequency wavelet coefficients to the time domain:}
\par This results in the benchmarked series $\tilde{Y}^{H}_{T}$:
\begin{align*}
\tilde{Y}^{H}_{T}&=\left(W^{H}\right)^{'}\hat{y}^{H} \\
&=\underbrace{\left(W^{H}\right)^{'}
\left[
\begin{array}{cc}
I_{m} & 0_{m,n-m,} \\
0_{n-m,m} & T_{n-m,n-m}(\tilde{y}^{H,H})
\end{array}\right]
\left[
\begin{array}{ccc}
0 & 0 & \frac{1}{c}I \\
0 & I & 0
\end{array}\right]
\left[
\begin{array}{cc}
W^{H} & 0 \\
0 & W^{L}
\end{array}\right]}_{\text{A=Elementary Wavelet Benchmarking and Thresholding Matrix}}
\left[
\begin{array}{c}
\check{Y}^{H}_{O}\\
Y^{L}_{O}
\end{array}\right]
\end{align*}
Matrix $A$ expresses the overall benchmarking process in a form consistent with equation \ref{benchmarkingsolutiongeneral}.
\par
\textbf{if} Seasonality is present \textbf{then}
\par Reintroduce the seasonal component $\hat{\gamma}$
\begin{center}
$\hat{Y}^{H}_{T}=\tilde{Y}^{H}_{T}+\hat{\gamma}$
\end{center}
\textbf{else}
\begin{center}
Set $\hat{Y}^{H}_{T}=\tilde{Y}^{H}_{T}$
\end{center}

\section{Data Analysis}\label{dataanalysissection}
We now consider the application of wavelet benchmarking to simulated data and an ONS data set. The advantages of a wavelet approach to benchmarking discussed in previous sections are supported by diagnostic measures of performance. Since simulated time series are additive, additive methods of benchmarking have been used. However, analogous results hold for multiplicative time series.

\subsection{Revision Metric for Benchmarking}
Subsequent sections assess benchmarking methods using average MSE and a revision metric. The average MSE metric assesses the performance of simulations but real data sets require an alternative metric since the true high frequency series is unobserved. As mentioned earlier, since published economic data impacts decisions made by policy makers, producing a stable benchmarked series is important. Therefore when current data sets are revised or new data becomes available adjustments to a benchmarked series should be minor. In particular the impact upon latter regions of the benchmarked series is most important since these points describe most recent economic conditions. The following metric measures the sensitivity of the latter regions of a benchmarked series when the observed high and low frequency series are adjusted.

Consider quarterly to annual GDP benchmarking; the series $\left\lbrace Y^{Q}_{O,t} \right\rbrace_{t=1,\ldots n}$ and $\left\lbrace Y^{A}_{O,t}\right\rbrace_{t=1,\ldots,m}$ are observed with corresponding benchmarked series $\left\lbrace \hat{Y}^{Q}_{T,t}\right\rbrace_{t=1,\ldots,n}$. When new data becomes available the new benchmarked series $\left\lbrace \tilde{Y}_{T,t}^{Q} \right\rbrace_{t=1,\ldots,l}$ is observed with $l\geq n$. A metric focusing on the last year of common benchmarked data is used. It measures the discrepancy between the last four quarters of overlapping time points \footnote{Equation \ref{Revisionmetricmean} provides larger metric readings for upward movements of the benchmarked series compared to downward movements. However since changes in benchmarked series are relatively small such differences are negligible.}:
\begin{align}
\label{Revisionmetricmean}&\text{Metric} =   100\times \frac{1}{4}\sum\limits_{t=n-3}^{n}\left(\left\lvert 1-\frac{\tilde{Y}^{Q}_{t}}{\hat{Y}^{Q}_{t}}\right\rvert\right)
\end{align}
Suppose $p$ years of additional data becomes available, so $\left\lbrace Y^{Q}_{O,t}\right\rbrace_{t=1,\ldots ,n+4p}$ and $\left\lbrace Y_{O,t}^{A}\right\rbrace_{t=1,\ldots,m+p}$ are now observed.
Consequently, we construnct $p$ new benchmarked series $\left\lbrace \tilde{Y}^{Q}_{O,t}\right\rbrace_{t=1,\ldots,n+4},$ $\ldots$, $\left\lbrace \tilde{Y}^{Q}_{O,t}\right\rbrace_{t=1,\ldots,n+4p}$. We compare each of these new benchmarked series and the original benchmarked series $\left\lbrace \hat{Y}_{T,t}^{Q} \right\rbrace_{t=1,\ldots,n}$.
In subsequent sections, the mean of these differences for a suitably chosen $p$, which will depend on the data length, will be referred to as the revision metric. In particular, it should be noted that this metric will be zero for both the original series and elementary wavelet benchmarking. In this case, additional data has no effect on the estimated high frequency series at earlier time points.

\subsection{Dyadic Quarterly and Annual Data}\label{dyadicquarterlyandannualdata}
Section \ref{applicationelementarwaveletbenchmarking} applied different benchmarking methods to 500 simulations; this section expands elementary wavelet benchmarking by introducing thresholding (Section \ref{thresholding}). Equation \ref{benchmarkedseries} decomposed the benchmarked quarterly series into a non noisy ($Y^{A,Q}_{T,t}$) and noisy ($R^{J+1,J+2}_{Y^{Q}_{O,t}}$) component:
\begin{align*}
\hat{Y}^{Q}_{T,t}=\underbrace{\frac{w^{A}(-1,0)}{2}\varphi_{-1,1}^{Q}(t)+\sum\limits_{j=0}^{J}\sum\limits_{k=0}^{k_{j}}\frac{w^{A}(j,k)}{2}\varphi^{Q}_{j,k}(t)}_{Y^{A,Q}_{T,t}}+\underbrace{\sum\limits_{j=J+1}^{J+2}\sum\limits_{k=0}^{k_{j}}w^{Q}(j,k)\varphi_{j,k}(t)}_{R^{J+1,J+2}_{Y^{Q}_{O,t}}}
\end{align*}
As mentioned previously, thresholding the noisy component should produce a more reliable series. However the structural form of the quarterly time series needs to be considered. Its seasonal component exists primarily on the high frequency regions of the wavelet domain. Thresholding has a tendency to interpret such subtle and localised features as noise; consequently thresholding inadvertently removes the seasonal component.

Removing the seasonal component prior to benchmarking/thresholding and reintroducing it afterwards offers one solution, as seen in Section \ref{alternativeseasonalmodelbenchmarking}. Therefore we used wavelet benchmarking with seasonal adjustment for analysis of the simulations.

The 500 simulations from Section \ref{applicationelementarwaveletbenchmarking} are reexamined. Average MSE and revision metric values (with $p=4$, corresponding to four additional years of data being available) for the different benchmarking methods are summarised in Table \ref{500simulationsdyadicquarterlyannualreexamined}\footnote{When additional data is introduced, it should be noted that data sets are no longer dyadic. Hence a traditional Haar basis is no longer appropriate to transform the data from the time to wavelet domain. Therefore an UB Haar basis was used.}. Clearly wavelet benchmarking outperforms all previous methods discussed so far; this is illustrated by its average MSE values being lower than the other benchmarking methods corresponding values. In terms of revisions, elementary wavelet benchmarking produces a benchmarked series which is not revised when new data becomes available. The revision metric value also implies that wavelet benchmarking outperforms currently used methods in terms of producing a stable benchmarked series (the same results were also qualitively found for other values of $p$, data not shown).
\begin{table}[H]
\begin{center}
\begin{tabular}{|c|c|c|}
\hline
Series Type & Average MSE & Revision Metric \\
\hline
Original &  2419.84    & 0.00  \\
Denton 1 &  1208.75  & 9.37   \\
Denton 2 &  1252.84  & 19.59   \\
Dagum and Cholette & 1203.51   & 16.48  \\
Elementary Wavelet Benchmarking & 1253.77  & 0.00   \\
Wavelet Benchmarking & 698.13  & 2.71 \\
\hline
\end{tabular}
\caption{Average MSE and metric values of different benchmarking methods corresponding to 500 dyadic quarterly and annual simulated series.}
\label{500simulationsdyadicquarterlyannualreexamined}
\end{center}
\end{table}

\subsection{Comparison to Current Methods}\label{nondyadicmonthlyquarterlydataexample1}
Simulated data from Section \ref{dyadicquarterlyandannualdata} relied upon the unrealistic assumption of both data sets having dyadic length. This assumption can be relaxed and now non dyadic monthly and quarterly data sets are analysed. Furthermore, the monthly series has a periodicity of three, resulting in a non dyadic relationship between these two data sets. As in Section \ref{dyadicquarterlyandannualdata} the model specified by equations \ref{STS1a}-\ref{STS7a} is used to generate the high frequency monthly data. Initialisation and parameter values used in simulations can be found in the Table \ref{parametervalues2}.

\begin{figure}[H]
\begin{center}
\includegraphics[width=\textwidth, height=10cm]{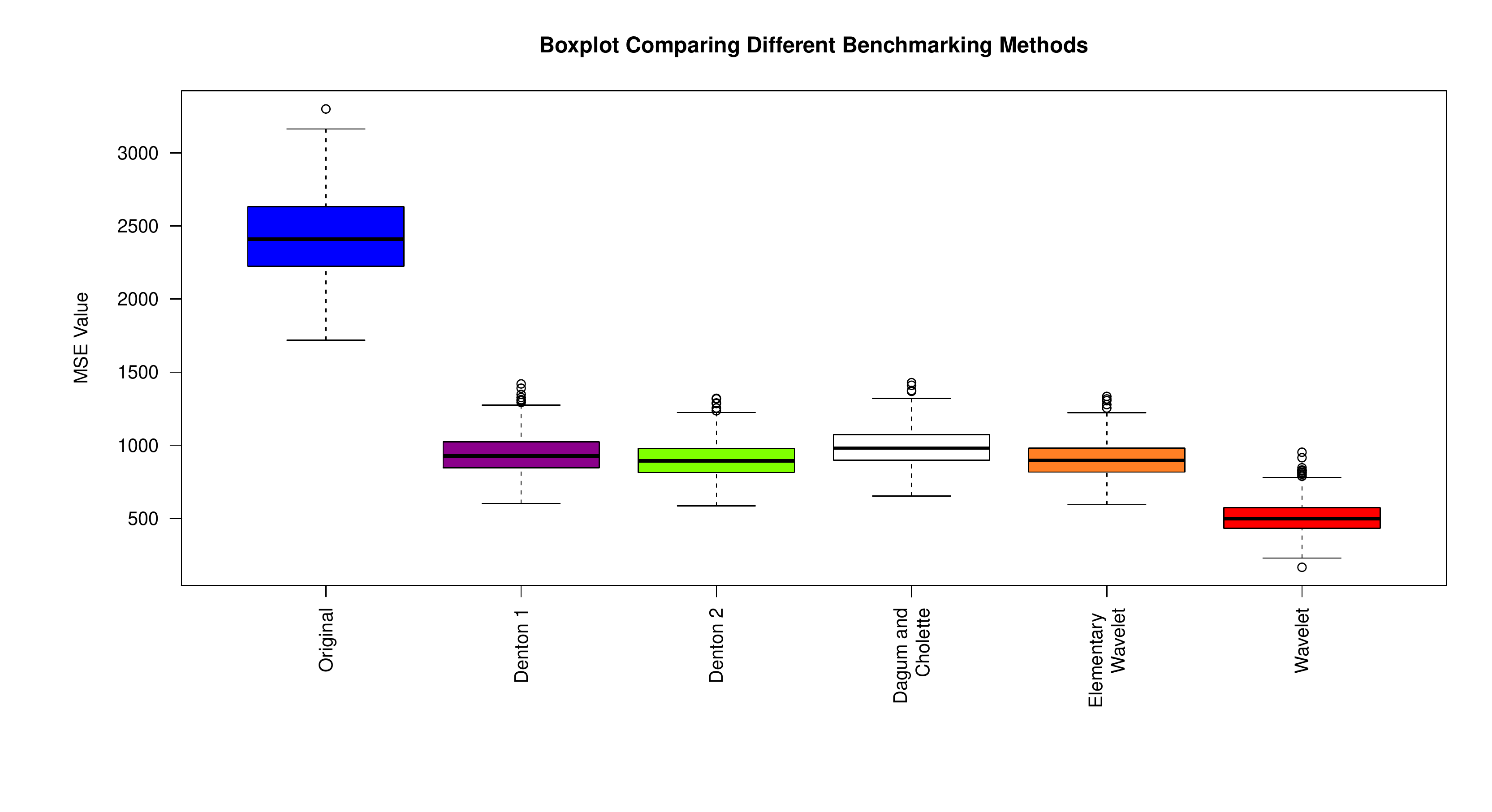}
\caption{Box plot comparing MSE values of different benchmarking methods for 500 simulations generated from non dyadic monthly and quarterly data sets.}
\label{msevaluesnondyadicmonthlyquarterlydataboxplot}
\end{center}
\end{figure}

Once again 500 simulations were generated \footnote{One simulation was removed since it caused a large distortion in the metric values resulting in an inaccurate comparison of the different benchmarking methods.}. Average MSE and metric values ($p=4$) for various benchmarking methods are recorded in Table \ref{nondyadicmonthlyandquarterlydataaveragemsevaluestable}. Figure \ref{msevaluesnondyadicmonthlyquarterlydataboxplot} shows a box plot comparing MSE values of the observed series to the benchmarked series.
\begin{table}[H]
\begin{center}
\begin{tabular}{|c|c|c|c|}
\hline
Series Type & Average MSE  & Revision Metric \\
\hline
Original (Noisy) & 2423.91  & 0.00 \\
Denton 1 & 904.11 & 11.71  \\
Denton 2 & 939.85  &  12.83  \\
Dagum and Cholette & 902.08 & 10.08 \\
Elementary Wavelet Benchmarking & 987.77  &0.00 \\
Wavelet Benchmarking & 506.81 & 3.56  \\
\hline
\end{tabular}
\caption{Average MSE and metric values comparing the original series to various benchmarking methods. 500 simulations from non dyadic monthly and quarterly data sets were generated.}
\label{nondyadicmonthlyandquarterlydataaveragemsevaluestable}
\end{center}
\end{table}
Results from Table \ref{nondyadicmonthlyandquarterlydataaveragemsevaluestable} and Figure \ref{msevaluesnondyadicmonthlyquarterlydataboxplot} are consistent with results from Section \ref{dyadicquarterlyandannualdata}. Elementary wavelet benchmarking peforms similarly to currently used benchmarking methods with improvements being offered using wavelet benchmarking. As would be expected wavelet benchmarking outperforms elementary wavelet benchmarking in terms of MSE for each of the 500 simulations. In all but one of the 500 simulations wavelet benchmarking outperformed both Denton and Dagum and Cholette in terms of MSE. In the one simulation wavelet benchmarking failed to outperform currently used methods the difference in MSE was negligible. The revision metric once again implies wavelet benchmarking produces a more stable benchmarked series in terms of revisions compared to currently used methods. Such evidence suggests wavelet benchmarking significantly outperforms currently used methods implemented by NSIs.

\subsection{Comparison to Current Methods (Shorter Series)}\label{comparisoncommonlyusedmethods2}
Previous examples used simulated time series with longs lengths not typically seen in time series published by NSIs. In reality time series being analysed have smaller lengths. Hence the performance of wavelet benchmarking in this setting is of interest. The same structural time series model defined by equations \ref{STS1a}-\ref{STS7a} was used to generate data. Parameter and initialisation values and can be found in Table \ref{parametervalues3}. The quarterly and monthly time series considered have respective lengths of $10$, $30$.

\begin{figure}[H]
\begin{center}
\includegraphics[width=\textwidth, height=10cm]{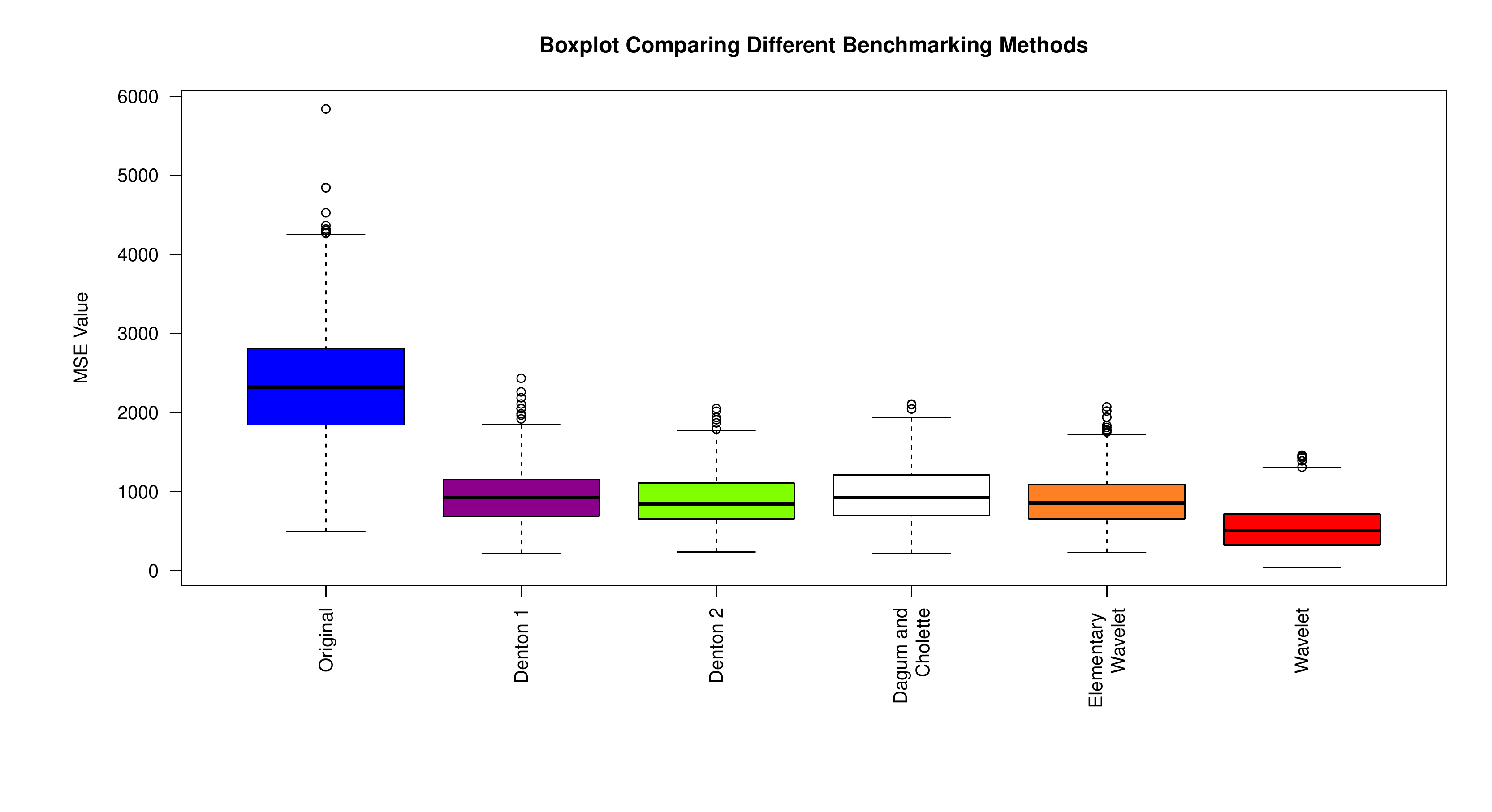}
\caption{Boxplot Comparing Denton, Dagum and Cholette and Wavelet Benchmarking.
500 simulations were generated from non dyadic monthly and quarterly series.}
\label{comparisondifferentbenchmarkingmethodexample41}
\end{center}
\end{figure}

Once again 500 simulations were generated \footnote{Two simulations were removed from the data sets since extreme wavelet benchmarking estimates were produced. This results from the input of poor initialisation values during the MLE procedure, and is easily seen as part of any reasonable quality control.} with results summarised in Table \ref{comparisondifferentbenchmarkingmethodexample3} and Figure \ref{comparisondifferentbenchmarkingmethodexample41}.
\begin{table}[H]
\begin{center}
\begin{tabular}{|c|c|c|c|}
\hline
Series Type & Average MSE & Revision Metric \\
\hline
Original (Noisy) & 2410.47  & 0.00 \\
Denton 1 & 921.85 &  23.47 \\
Denton 2 & 979.42 & 37.43 \\
Dagum and Cholette & 914.92  & 20.31 \\
Elementary Wavelet &  994.16 & 0.00 \\
Wavelet Benchmarking & 562.61  & 18.34 \\
\hline
\end{tabular}
\caption{Average MSE and metric values for different benchmarked series corresponding to
500 non dyadic monthly and quarterly simulated series.}
\label{comparisondifferentbenchmarkingmethodexample3}
\end{center}
\end{table}
As expected wavelet benchmarking outperforms both Denton and Dagum and Cholette benchmarking; however improvements from wavelet benchmarking are reduced. This is reflected by comparing average MSE values recorded in Table \ref{nondyadicmonthlyandquarterlydataaveragemsevaluestable} and Table \ref{comparisondifferentbenchmarkingmethodexample3}. The percentage reduction in
average MSE using wavelet benchmarking is greater in Table \ref{nondyadicmonthlyandquarterlydataaveragemsevaluestable} (long time series) compared to its corresponding value in Table \ref{comparisondifferentbenchmarkingmethodexample3} (short time series). For shorter time series the revision metric (here with $p=2$, given the short length of the series) shows that wavelet benchmarking produce more stable benchmarked series compared to currently used methods.

\subsection{Official ONS Data}
\begin{figure}[H]
\begin{center}
\includegraphics[width=\textwidth, height=12cm]{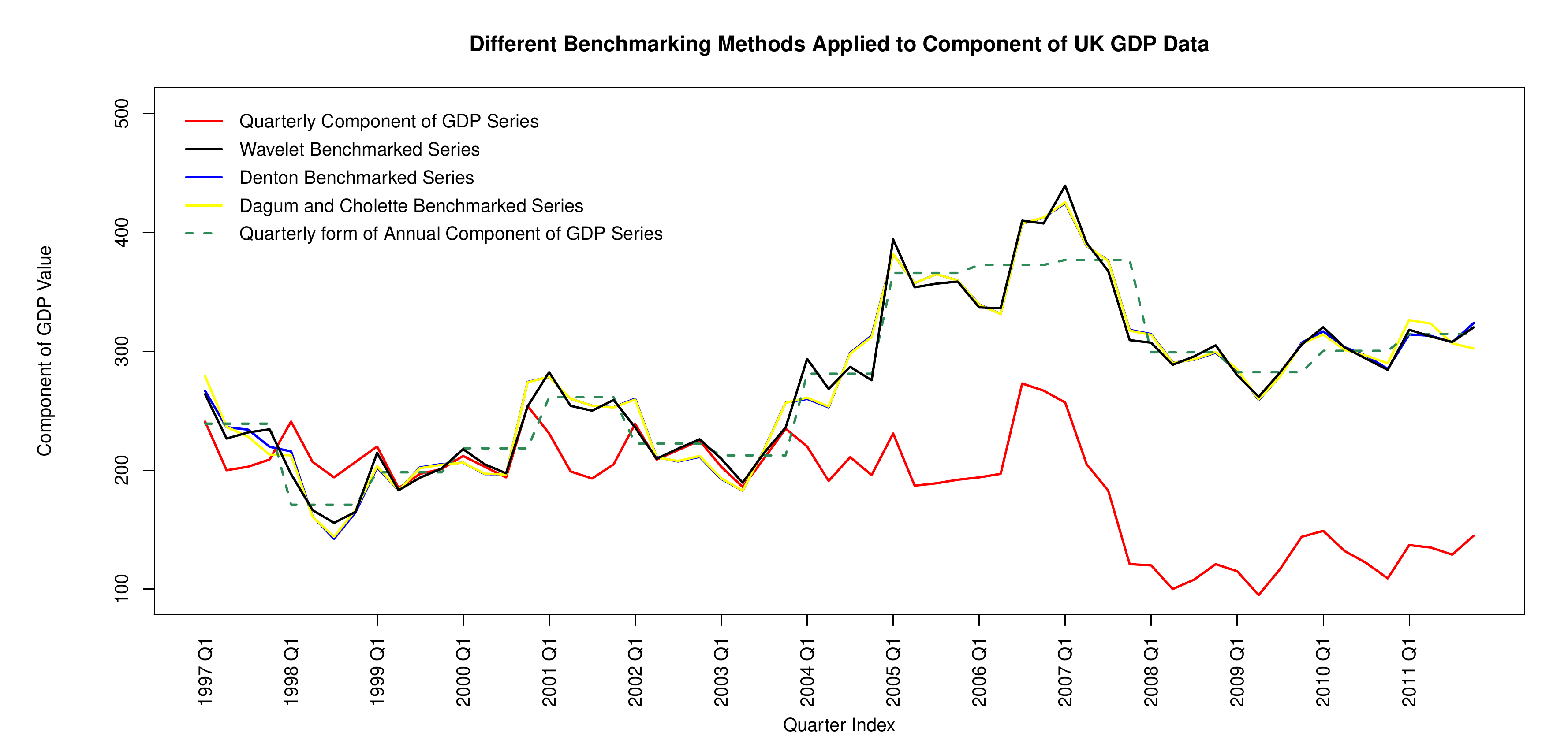}
\caption{Application of Various Benchmarking Methods to Component of UK GDP Data.}
\label{ONSrealdatabenchmakringexample}
\end{center}
\end{figure}

The following section investigates the application of various benchmarking methods to official ONS data. Data from UK national accounts is analysed; in particular one component of GDP data is considered. For confidentiality reasons this component can not be named. Figure \ref{ONSrealdatabenchmakringexample} shows the results of applying quarterly to annual benchmarking to this one component of GDP data.
In Figure \ref{ONSrealdatabenchmakringexample} the Denton and Dagum and Cholette versions of the benchmarked series perform similarly. The output of wavelet benchmarking is similar to currently used methods; however in some time periods wavelet benchmarking performs better at preserving movements in the observed quarterly series. One such time period is from 2004 Q1 to 2005 Q1. This is due to the localised nature of a wavelet benchmarking solution.
\par
In the time period 2007 Q1 to 2008 Q1 the observed quarterly time series seems to exhibit a structural break. This structural break most likely is a result of the economic recession which began in 2007. By creating a wavelet basis which considers the structure of the observed time series, wavelet benchmarking has the ability to offer further improvements in terms of ensuring movements in the quarterly time series are persevered. In this paper wavelet bases are solely determined by the length of observed time series. Future work could incorporate the structure of these time series during the selection of wavelet bases.
\par
Table \ref{comparisondifferentbenchmarkingmethodexample4} records the metric values for the different benchmarking methods. Wavelet benchmarking produces a stable benchmarked series and on whole performs similarly to currently used methods. For this example the maximum lag length $p=3$ was used, since this corresponds to $20\%$ of the length of the observed series. However, results were not qualitatively different for smaller values of $p$.
\begin{table}[H]
\begin{center}
\begin{tabular}{|c|c|}
\hline
Series Type & Revision Metric \\
\hline
Original (Noisy)  &  0.00 \\
Denton 1 & 0.61 \\
Denton 2  &  0.35 \\
Dagum and Cholette  & 1.95 \\
Elementary Wavelet  & 0.00\\
Wavelet Benchmarking   & 0.87\\
\hline
\end{tabular}
\caption{Metric values for different benchmarked series corresponding to
official ONS data.}
\label{comparisondifferentbenchmarkingmethodexample4}
\end{center}
\end{table}

\section{Discussion}\label{Conclusion}
Benchmarking is a problem frequently encountered by NSIs; this paper provided an introduction to wavelet based solutions. Wavelet based benchmarking consists of a non parametric and a parametric step. The first step involved introducing non noisy information from the benchmark series into the noisy observed high frequency series via the wavelet domain. Afterwards high frequency wavelet coefficients were thresholded to remove any remaining noise. However the structural form of time series being analysed had to be considered; in particular the seasonal component is often incorrectly identified as noise and inadvertently removed. Consequently periodic structural time series models were used to seasonally adjust the high frequency series while ensuring the benchmark constraint was satisfied. After thresholding the seasonally adjusted high frequency series the estimated seasonal component was reintroduced to form the final benchmarked series.

To illustrate wavelet benchmarking both simulated and real data sets were analysed. Simulation studies showed that wavelet benchmarking outperformed currently used methods.

By forcing the benchmarked series to be consistent with the benchmark series there
is an implicit and unrealistic assumption that the benchmark series is not contaminated with
noise. This assumption can be relaxed; both high and low frequency processes can
be treated as noisy. Benchmarking can now be described as optimally combining
both high and low frequency processes to create a benchmarked series. It can
also be extended to situations where multiple constraints must be satisfied. One
such example occurs when a time series is classified according to periodicity
and geographical location. Benchmark constraints need to be satisfied on both
individual and aggregate levels; wavelet benchmarking could facilitate this too.

The following four areas could be considered to extend work on wavelet benchmarking. Firstly seasonal adjustment could be performed in the wavelet domain, thus allowing the entire benchmarking problem to be considered in the wavelet domain. Secondly, while this paper considered binding benchmarking (the low frequency series is assumed to be non noisy), this assumption could be dropped and wavelet benchmarking in the setting of observing noisy low and high frequency series could be considered. Thirdly the selection of wavelet bases needs to be considered in greater detail. This paper constructed such bases based on the length of observed time series. While a reasonable starting point for an introduction to wavelet benchmarking, bases which incorporate the structure of observed time series could be used in future work. Finally the ONS performs benchmarking on a large number of time series and therefore would require a method of wavelet benchmarking which can be used in a mass production setting.

\bibliographystyle{abbrv}
\bibliography{dkbook}

\begin{thebibliography}{10}

\bibitem{BinmoreDavies(2002)}
K.~Binmore and J.~Davies.
\newblock {\em Calculus: Concepts and Methods}.
\newblock Cambridge University Press, Cambridge, 2002.

\bibitem{BrownPS2012}
G.~Brown, N.~Parkin, and N.~Stuttard.
\newblock A review of benchmarking methods.
\newblock In {\em Seventeenth GSS Conference on Methodology}, 2012.

\bibitem{DagumCholetteBook(2006)}
P.~Cholette and E.~Dagum.
\newblock {\em Benchmarking, Temporal Distribution, and Reconciliation Methods
  for Time Series}.
\newblock Springer, New York, 2006.

\bibitem{DagumCholette(94)}
P.~Cholette and E.~B. Dagum.
\newblock Benchmarking time series with autocorrelated survey errors.
\newblock {\em International Statistical Review}, 62:365--377, 1994.

\bibitem{Daubechies(92)}
I.~Daubechies.
\newblock {\em Ten Lectures on Wavelets}.
\newblock Society for Industrial and Applied Mathematics, 1992.

\bibitem{Denton(71)}
F.~T. Denton.
\newblock Adjustment of monthly or quarterly series to annual totals: An
  approach based on quadratic minimization.
\newblock {\em J. American Statistical Association}, 66:99--102, 1971.

\bibitem{Donoho(94)}
D.~L. Donoho and I.~M.~. Johnstone.
\newblock Ideal spatial adaptation by wavelet shrinkage.
\newblock {\em Biometrica}, 81:425--455, 1994.

\bibitem{DurbinKoopman(2001)}
J.~Durbin and S.~J. Koopman.
\newblock {\em Time Series Analysis by State Space Methods}.
\newblock Oxford University Press, Oxford, 2001.

\bibitem{DurbinQuenneville(97)}
J.~Durbin and B.~Quenneville.
\newblock Benchmarking by state space models.
\newblock {\em International Statistical Review}, 65:23--48, 1997.

\bibitem{findley2005}
D.~F. Findley.
\newblock Some recent developments and directions in seasonal adjustment.
\newblock {\em J. Official Statistics}, 21(2):343--365, 2005.

\bibitem{Fryzlewicz(2007)}
P.~Fryzlewicz.
\newblock Unbalanced haar technique for nonparametric function estimation.
\newblock {\em J. American Statistical Association}, 102:1318--1327, 2007.

\bibitem{Nason(2008)}
G.~Nason.
\newblock {\em Wavelet Methods in Statistics with R}.
\newblock Springer Publishing Company, Incorporated, 1 edition, 2008.

\bibitem{Nason(99)}
G.~Nason, R.~von Sachs, and G.~Kroisandt.
\newblock Wavelet processes and adaptive estimation of the evolutionary wavelet
  spectrum.
\newblock {\em J. Royal Statistical Society B}, 61:63--84, 1999.

\bibitem{PercivalWalden(2000)}
D.~B. Percival and A.~T. Walden.
\newblock {\em Wavelet Methods for Time Series Analysis}.
\newblock Cambridge University Press, Cambridge, 200.

\bibitem{Percival(95)}
D.~P. Percival.
\newblock On estimation of the wavelet variance.
\newblock {\em Biometrika}, 82:619--631, 1995.

\bibitem{Rsoftware}
{R Development Core Team}.
\newblock {\em R: A Language and Environment for Statistical Computing}.
\newblock R Foundation for Statistical Computing, Vienna, Austria, 2008.
\newblock {ISBN} 3-900051-07-0.

\bibitem{Tripodis(2004)}
Y.~Tripodis and J.~Penzer.
\newblock Periodic time series models: a structural approach.
\newblock Technical report, London School of Economics, 2004.

\bibitem{Vidakovic(99)}
B.~Vidakovic.
\newblock {\em Statistical Modeling by Wavelets}.
\newblock Wiley-Blackwell, New York, 1999.

\end{thebibliography}
\newpage
\section{Appendix}
\subsection{Simulation Methodology}\label{appendixSTSmodel}
The following section describes how simulated time series were generated. The
model below generates the unobserved true high frequency data points.
\begin{align}
&Y^{H}_{T,t}=\mu_{t}+\gamma_{t} \label{STS1a} \\
&\mu_{t}=\mu_{t-1}+\upsilon_{t}+\varphi_{t}, \hspace{2mm} \varphi_{t}\sim
N(0,\sigma^{2}_{\varphi})\label{STS3a}\\
&\upsilon_{t}=\upsilon_{t-1}+\zeta_{t}, \hspace{2mm} \zeta_{t}\sim
N(0,\sigma^{2}_{\zeta}) \label{STS4a} \\
&\gamma_{t}=-\sum\limits_{i=1}^{k-1}\gamma_{t-i} +\omega_{t}, \hspace{2mm}
\omega_{t}\sim N(0,\sigma^{2}_{\omega})\label{STS2a}
\end{align}
The observed non noisy low frequency time series is obtained using:
\begin{align}
Y^{L}_{O,t}=\sum\limits_{i=kt-(k-1)}^{kt}Y^{H}_{T,t} \label{STS5a}
\end{align}
An ARMA(1,1) process is used throughout the paper to generate disturbance terms.
This results in the following observed high frequency series.
\begin{align}
&Y^{H}_{O,t}=Y^{H}_{T,t}+\epsilon_{t}, \hspace{2mm} \epsilon_{t}\sim ARMA(1,1)
\label{STS6a} \\
&\epsilon_{t}=\phi\epsilon_{t-1}+\theta\tau_{t},
\hspace{2mm}\tau_{t}\sim\mathbb{N}(0,\sigma^{2}_{\tau}),
\hspace{2mm}|\phi|,|\theta| <1 \label{STS7a}
\end{align}
Initialisation values are required to begin the simulation. The values
$\mu_{1},\upsilon_{1},\gamma_{1},\ldots,\gamma_{k-1}$ are generated
independently from a zero mean Gaussian process with respective variances
$\sigma_{\mu_{1}}^{2},\sigma_{\upsilon_{1}}^{2},\sigma_{\gamma_{1}}^{2},\ldots,
\sigma_{\gamma_{k-1}}^{2}$

To ensure simulations can be reproduced the set.seed() \cite{Rsoftware} function is used to
generate pseudo random numbers. For the slope, trend, seasonal componenets the
following pseudo random numbers are used respectively; set.seed(simulation
number $\times$ time series number), set.seed($2\times$ simulation number
$\times$ time series number), set.seed($3\times$ simulation number $\times$ time
series number). The term simulation number identifies the current simulation
being generated, while time series number corresponds to the time point in that
current simulation.

\subsection{Simulations for Elementary Wavelet Benchmarking}\label{simulationvalueelementarywaveletbenchmarking}
Initialisation and parameter values used to generate simulations for time series
analysed in Section \ref{applicationelementarwaveletbenchmarking} and Section \ref{dyadicquarterlyandannualdata} are
summarised in Table \ref{parametervalues1}.
\begin{table}[H]
\begin{center}
\begin{tabular}{|c|c|c|c|c|c|c|c|c|c|c|c|c|c|c|}
\hline
$\sigma_{\mu_{1}}$ & $\sigma_{\upsilon_{1}}$ & $\sigma_{\gamma_{1}}$ &
$\sigma_{\gamma_{2}}$ & $\sigma_{\gamma_{3}}$ & $\phi$ & $\theta$ &
$\sigma_{\varphi}$ & $\sigma_{\zeta}$ & $\sigma_{\omega}$ & $m$ & $n$ & $k$ & $p$  \\
\hline
$1$ & $1$ & $1$ & $1$ & $1$ & $0.2$ & $0.5$ & $1$ & $0.25$ & $3$ & $2^{6}$ &
$2^{8}$ & $4$ & $4$ \\
\hline
\end{tabular}
\caption{Parameter and Initialisation values used to generate simulations in
Section \ref{applicationelementarwaveletbenchmarking} and Section \ref{dyadicquarterlyandannualdata}}
\label{parametervalues1}
\end{center}
\end{table}

\subsection{Simulations for Data Section Analysis}
\subsubsection{Comparison to Current
Methods}\label{appendixnondyadicmonthlyandquarterlydata}
Parameter and initialisation values used to generate time series analysed in
Section \ref{nondyadicmonthlyquarterlydataexample1} are recorded below in Table
\ref{parametervalues2}.
\begin{table}[H]
\begin{center}
\begin{tabular}{|c|c|c|c|c|c|c|c|c|c|c|c|c|c|}
\hline
$\sigma_{\mu_{1}}$ & $\sigma_{\upsilon_{1}}$ & $\sigma_{\gamma_{1}}$ &
$\sigma_{\gamma_{2}}$ & $\phi$ & $\theta$ & $\sigma_{\varphi}$ &
$\sigma_{\zeta}$ & $\sigma_{\omega}$ & $m$ & $n$ & $k$ & $p$ \\
\hline
$1$ & $1$ & $1$ & $1$ & $0.2$ & $0.5$ & $1$ & $0.25$ & $3$ & $70$ & $210$ & $3$ & $4$
\\
\hline
\end{tabular}
\caption{Parameter and Initialisation values used to generate simulations in
Section \ref{nondyadicmonthlyquarterlydataexample1}.}
\label{parametervalues2}
\end{center}
\end{table}

\subsubsection{Comparison to Current Methods (Shorter
Series)}\label{appendixnondyadicmonthlyandquarterlydata}
Parameter and initialisation values used to generate time series analysed in
Section \ref{comparisoncommonlyusedmethods2} are recorded below in Table
\ref{parametervalues3}.
\begin{table}[H]
\begin{center}
\begin{tabular}{|c|c|c|c|c|c|c|c|c|c|c|c|c|c|}
\hline
$\sigma_{\mu_{1}}$ & $\sigma_{\upsilon_{1}}$ & $\sigma_{\gamma_{1}}$ &
$\sigma_{\gamma_{2}}$ & $\phi$ & $\theta$ & $\sigma_{\varphi}$ &
$\sigma_{\zeta}$ & $\sigma_{\omega}$ & $m$ & $n$ & $k$ & $p$  \\
\hline
$1$ & $1$ & $1$ & $1$ & $0.2$ & $0.5$ & $1$ & $0.25$ & $3$ & $10$ & $30$ & $3$ & $2$
\\
\hline
\end{tabular}
\caption{Parameter and Initialisation values used to generate simulations in
Section \ref{comparisoncommonlyusedmethods2}.}
\label{parametervalues3}
\end{center}
\end{table}

\end{document}